\documentclass[runningheads]{cl2emult}

\usepackage{makeidx}  
\usepackage{graphicx} 
\usepackage{subeqnar} 
\usepackage{multicol} 
\usepackage{cropmark} 
\usepackage{physbb}   
\makeindex            

\usepackage{amssymb}
\def\greaterthansquiggle{\raise.3ex\hbox{$>$\kern-.75em\lower1ex\hbox{$\sim$}}}
\def\lessthansquiggle{\raise.3ex\hbox{$<$\kern-.75em\lower1ex\hbox{$\sim$}}}
\newcommand{\beq}{\begin{equation}}
\newcommand{\eeq}{\end{equation}}
\newcommand{\beqa}{\begin{eqnarray}}
\newcommand{\eeqa}{\end{eqnarray}}
\newcommand{\beqan}{\begin{eqnarray*}}
\newcommand{\eeqan}{\end{eqnarray*}}
\newcommand{\ba}{\begin{array}}
\newcommand{\ea}{\end{array}}
\newcommand{\no}{\nonumber}

\newcommand{\ra}{\rightarrow}
\newcommand{\Ra}{\Rightarrow}
\newcommand{\ve}{\varepsilon}
\newcommand{\vp}{\varphi}

\newcommand{\wt}{\widetilde}

\newcommand{\C}{{\cal C}}

\newcommand{\cL}{{\cal L}}

\newcommand{\cP}{{\cal P}}

\newcommand{\R}{{\cal R}}

\newcommand{\st}{\stackrel}

\def\nz{\ifmmode {I\hskip -3pt N} \else {\hbox {$I\hskip -3pt N$}}\fi}
\def\zz{\ifmmode {Z\hskip -4.8pt Z} \else
       {\hbox {$Z\hskip -4.8pt Z$}}\fi}
\def\qz{\ifmmode {Q\hskip -5.0pt\vrule height6.0pt depth 0pt
       \hskip 6pt} \else {\hbox
       {$Q\hskip -5.0pt\vrule height6.0pt depth 0pt\hskip 6pt$}}\fi}
\def\rz{\ifmmode {I\hskip -3pt R} \else {\hbox {$I\hskip -3pt R$}}\fi}
\def\cz{\ifmmode {C\hskip -4.8pt\vrule height5.8pt\hskip 6.3pt} \else
       {\hbox {$C\hskip -4.8pt\vrule height5.8pt\hskip 6.3pt$}}\fi}

\begin{document}
\title*{Time-Independent Gravitational Fields}
\toctitle{Time-Independent Gravitational Fields}
%
%
\titlerunning{Time-Independent Gravitational Fields}
%
\author{R. Beig\inst{1}
\and B. Schmidt\inst{2}
}
\authorrunning{R.~Beig et al.}
%
%
\institute{Institut f\"ur Theoretische Physik, Universit\"at Wien, 
Boltzmanngasse 5, \\ A--1090 Wien, \"Osterreich
\and Max-Planck-Institut f\"ur Gravitationsphysik \\
Albert-Einstein-Institut \\ M\"uhlenweg 1, D-14476 Golm b. Potsdam,
Deutschland }
\maketitle              

\makeatletter                  
\@addtoreset{equation}{section}
\renewcommand{\theequation}{\arabic{section}.\arabic{equation}}%
\renewcommand{\thesubequation}{\arabic{section}.%
                               \arabic{equation}\alph{eqsubcnt}}%
\makeatother                   

\section{Introduction}
In this article we want to describe what is known about time independent
spacetimes from a global point of view. The physical situations we want to treat
are isolated bodies at rest or in uniform rotation in an otherwise empty
universe. In such cases one expects the gravitational field to have no
``independent degrees of freedom''. Very loosely speaking, the spacetime
geometry should be uniquely determined by the matter content of the model
under consideration. In a similar way, for a given matter model (such as
that of a perfect fluid), there should be a one-to-one correspondence
between Newtonian solutions and general relativistic ones.

The plan of this paper is as follows. In Sect.~2.1 we collect the information
afforded on one hand by the Killing equation obeyed by the vector field
generating the stationary isometry and, on the other hand, that by the
Einstein field equations. 
Throughout this paper we assume this stationary isometry to be everywhere
timelike. Thus ergoregions are excluded.
We try as much as possible to write
the resulting equations in terms of objects intrinsic to the space
(henceforth simply called ``the quotient space'')
obtained by quotienting spacetime by the action of the stationary isometry.
Much of this is standard. But since none of the references known to us
meets our specific purposes, we give a self-contained treatment starting
from scratch. Since all the models we treat are axially symmetric, we 
add in Sect.~2.2 a second, axial Killing vector to the formalism of 
Sect.~2.1.

In Sect.~2.3 we first introduce new dependent variables for the vacuum
gravitational field, namely a conformally rescaled metric on the quotient
space and two potentials, using which the field equations have an
interpretation in terms of harmonic maps from the quotient space into
the Poincar\'e half plane. These potentials, originally due to Hansen,
are used in our treatment of asymptotics in Sects.~3.1,2. We then
formulate the boundary conditions at spatial infinity appropriate for
isolated systems and prove two basic theorems due to Lichnerowicz on
stationary solutions obeying these conditions. These theorems are
manifestations of the above-mentioned principle concerning the lack of
gravitational degrees of freedom. The first result, the ``staticity
theorem'', basically states that the gravitational field is static when
the matter is non-rotating. The second one, the ``vacuum theorem'',
states that spacetime is Minkowski when no matter is present. In 
Sect.~2.4 we restore $c$, the velocity of light, in the field equations
and show that these tend to the Newtonian ones as $c \to \infty$.

In Chapter~3 we  we study solutions only ``near infinity''.
(Note that by the Lichnerowicz vacuum theorem, such solutions can not be
 extended to all of ${\bf R^4}$ exept for flat spacetime.)
 In the Newtonian case
such solutions are known to have a convergent expansion in negative
powers of the radius where the coefficients are given by multipole
moments. The relativistic situation is slightly at variance with our 
statement at the beginning concerning the Newton--Einstein correspondence:
namely, there are now, corresponding to the presence of two potentials
rather than one, two infinite sequences of multipole moments, the
``mass moments'' which have a Newtonian analogue and the ``angular
momentum moments'' which do not. One may now study the two potentials
and the rescaled quotient space metric in increasing powers of $1/r$,
where $r$ is the radius corresponding to a specific coordinate gauge
on the quotient space which has to be readjusted at each order in
$1/r$.

The results one finds are sufficient for the existence of a chart in the 
one-point ``compactification'' of the quotient space (i.e. the union of
the quotient space and the point-at-infinity), in terms of which yet
another conformal rescaling of the 3-metric, together with a
corresponding rescaling of the two potentials, admit regular extensions
to the compactified space. As summarized in Sect.~3.2 one is then able
to find field equations for these ``unphysical'' variables which are
regular at the point-at-infinity and in addition can be turned into an
elliptic system. From this it follows that the unphysical quantities
are in fact analytic near infinity and this, in turn, implies
convergence for a suitable $1/r$-expansion ($r$ being a ``physical''
radius) for the original physical variables.
Furthermore the structure of the unphysical equations yields the result
that the (physical) spacetime metric is uniquely characterized by the two
sets of multipole moments. 

It is remarkable that stationary vacuum solutions satisfying rather
weak fall--off conditions at spatial infinity, by the very nature of the field 
equations, have to have a convergent multipole expansion. We believe
that the topic of far--field behaviour of time--independent gravitational fields 
 is by now reasonably well understood. The main open problem is to characterize an
 a priori given sequence of multipole moments for which the expansion converges.

In Chapter~4 we review global rotating solutions.
In Sect.~4.1 we outline a result due to Lindblom which shows that
stationary rotating spacetimes with a one-component fluid source with
phenomenological heat conduction and viscosity have to be axisymmetric.
In Sect.~4.2 we describe a theorem of Heilig which proves the existence of
axisymmetric, rigidly rotating perfect-fluid spacetimes with polytropic
equation of state, provided the parameters are sufficiently close to
ones for a nonrotating Newtonian solution. In Sect.~4.3 we present
the solution of Neugebauer and Meinel representing a rigidly rotating
infinitely thin disk of dust. In the final chapter we treat global
nonrotating solutions.

 In Sect.~5.1 we outline the essentials of a
relativistic theory of static elastic bodies. The remaining sections are devoted
to spherical symmetry. It has long been conjectured that nonrotating perfect fluids
are spherical whence Schwarzschild in their exterior region. In Sect.~5.2 
we discuss the present status of this conjecture. A proof exists when the 
allowed equations of state are limited by a certain inequality. While this 
inequality covers many cases of physical interest, the Newtonian situation
suggests that the conjecture is probably true without this restriction. 
In Sect.~5.3 we review spherically symmetric perfect fluid solutions. 
The final Sect.~5.4 gives a
 short description of self--gravitating Vlasov matter in the sperically 
symmetric case.

In the subject of time--independent gravitational field of isolated bodies
there are some topics we do not cover. We do not address the question of the
conjectured non--existence of solutions with more than one body. (M\"uller zum Hagen
\cite{MzH2B} has some results on this in the static case.) Furthermore we limit 
ourselves to ``standard matter'' sources. Thus Black Holes are excluded. 
(For this see the article of Maison in this volume.) We also could not cover the 
interesting case of soliton--like
 solutions for ``non--linear matter sources'', starting with the discovery of the
 Bartnik--McKinnon solutions of the Einstein--Yang Mills system (see Bizon
\cite{Biz}.)

\section*{Acknowledgements}
Part of this work was carried out at the Institute for Theoretical Physics
at Santa Barbara, where the authors took part in the program on "Classical
and Quantum Physics of Strong Gravitational Fields". We thank the ITP 
for its support and kind hospitality.
R.~Beig was in part supported by ``Fonds zur F\"orderung der 
wissenschaftlichen Forschung in \"Osterreich'', project P12626-PHY.

\section{Field Equations}
\subsection{Generalities}

Let $(M,g_{\mu\nu})$ be a 4-dimensional smooth connected manifold with
 Lorentz metric $g_{\mu\nu}$ of signature $(-+++)$. We assume
$M$ to be chronological, i.e. to admit no closed timelike curves.
Let $\xi^\mu$ be an everywhere timelike Killing vector field with
complete orbits. Thus we do not allow points where $\xi^\mu$ turns null,
i.e. we exclude horizons and ergospheres. It follows (see \cite{Ha}) that the
quotient of $M$ by the isometry group generated by $\xi^\mu$ is a
Hausdorff manifold $N$ and that $M$ is a principal ${\bf R}^1$-bundle
over $N$. Furthermore this bundle is trivial, i.e. $M$ is diffeomorphic
to ${\bf R}^1 \times N$. The fact that this diffeomorphism is non-natural
(whereas of course the projection $\pi$ mapping $M$ onto $N$ is) plays a
role in the formalism we shall now develop.

Let us introduce the differential geometric machinery necessary for writing
the stationary Einstein equations in a way naturally adapted to $\xi^\mu$.
As far as possible, we will be interested in quantities and equations 
intrinsic to $N$ (``dimensional reduction''). For a similar treatment
see the Appendix of \cite{GeS}. We define the fields $V$ and
$\omega_{\lambda\nu\lambda} = \omega_{[\mu\nu\lambda]}$ by
\beqa
V &:=& \xi_\mu \xi^\mu \Ra V < 0 \\
\omega_{\mu\nu\lambda} &:=& 3 \xi_{[\mu} \nabla_\nu \xi_{\lambda]} .
\eeqa
The 3-form $\omega_{\mu\nu\lambda}$ vanishes if and only if $\xi^\mu$ is
hypersurface orthogonal -- in which case $(M,g_{\mu\nu})$ is called static.
More important than $\omega_{\mu\nu\lambda}$ will be the 2-form 
$\sigma_{\mu\nu}$, given by
\beq
\sigma_{\mu\nu} := \omega_{\mu\nu\lambda} \xi^\lambda .
\eeq
Given $\xi^\mu$, the fields $\sigma_{\mu\nu}$ and $\omega_{\mu\nu\lambda}$ carry the
same information, since
\beq
\omega_{\mu\nu\lambda} = 3 V^{-1} \xi_{[\mu} \sigma_{\nu\lambda]}.
\eeq
Equ. (2.4) is obtained by expanding the identity 
$\xi_{[\mu} \omega_{\nu\lambda\rho]} = 0$, which follows from (2.2), and 
contracting with $\xi^\mu$. In a similar way we obtain the relations
\beqa
\omega_{\mu\nu\lambda} \omega^{\mu\nu\lambda} &=& 3V^{-1} \sigma_{\mu\nu}
\sigma^{\mu\nu} \\
\omega_{\mu\nu\lambda} \sigma^{\nu \lambda} &=& \frac{1}{3}
\omega_{\rho \nu \lambda} \omega^{\rho \nu \lambda} \xi_\mu.
\eeqa
We now invoke the Killing equation for $\xi^\mu$, i.e.
\beq
\cL_\xi g_{\mu\nu} = \nabla_\mu \xi_\nu + \nabla_\nu \xi_\mu = 0.
\eeq
Expanding $\omega_{\mu\nu\lambda}$ in terms of $\xi_\mu$, we easily see
that
\beq
\nabla_\mu \xi_\nu = V^{-1} [\sigma_{\mu\nu} + (\nabla_{[\mu} V)
\xi_{\nu]}],
\eeq
or, equivalently,
\beq
\sigma_{\mu\nu} = V^2 \nabla_{[\mu} (V^{-1} \xi_{\nu]}).
\eeq
In the static case we have $\sigma_{\mu\nu} = 0$, whence there exist
global cross sections given by $t =$~const, where $\xi_\mu = 
V \nabla_\mu t$. 

Equ. (2.9) implies that
\beq
\nabla_{[\mu} (V^{-2} \sigma_{\nu \lambda]}) = 0 .
\eeq
Clearly we have $\cL_\xi \tau = 0$, where $\tau$ is the 3-form given by
$\tau_{\mu\nu\lambda} = V^{-2} \omega_{\mu\nu\lambda}$. By (2.10) and
the identity $\cL_\xi = \xi \rfloor d \tau + d(\xi \rfloor \tau)$, this
implies $\xi \rfloor d\tau = 0$. Since $d\tau$ is a 4-form and $\xi^\mu$
is nowhere zero, we infer in 4 dimensions that $d\tau$ is zero, i.e.
\beq
\nabla_{[\mu} (V^{-2} \omega_{\nu\lambda\rho]}) = 0.
\eeq
Equ.'s (2.10,11) are integrability conditions for the Killing equations
(2.7) which are ``purely geometric'' in that they do not involve the
Ricci (whence: energy-momentum) tensor. Now recall the relation
\beq
\nabla_\mu \nabla_\nu \xi_\lambda = - R_{\nu\lambda\mu}{}^\rho \xi_\rho,
\eeq
which follows from (2.7) and its corollary
\beq
           g^{\nu\rho} \nabla_\nu \nabla_\rho \xi_\mu =
       - R_\mu{}^\nu \xi_\nu .
\eeq
From (2.2), (2.7) and (2.13) we find that
\beq
\nabla^\mu \omega_{\mu\nu\lambda} = 2 \xi_{[\nu} R_{\lambda ] \mu} \xi^\mu ,
\eeq
which, using (2.8), implies
\beq
\nabla^\mu (V^{-1} \sigma_{\mu\nu}) = 2V^{-1} \xi_{[\nu} R_{\lambda]\mu}
\xi^\mu \xi^\lambda - V^{-3} \sigma_{\mu\lambda} \sigma^{\mu\lambda} 
\xi_\nu ,
\eeq
where we have also used (2.6,7). Interpreting $G_{\mu\nu} = R_{\mu\nu} -
\frac{1}{2} g_{\mu\nu} R$ as the energy-momentum tensor of matter, the
r.h. side of Equ. (2.14) is zero iff the matter current, for an observer
at rest relative to $\xi^\mu$, is zero. In that case, and provided that
$M$ is simply connected, there exists a scalar field $\omega$, called
twist potential, such that
\beq
\omega_{\mu\nu\lambda} = \frac{1}{2} \ve_{\mu\nu\lambda}{}^\rho 
\nabla_\rho \omega ,
\eeq
and then (2.11) implies
\beq
\nabla^\mu (V^{-2} \nabla_\mu \omega) = 0 .
\eeq
Note that, by virtue of $\xi_{[\mu} \omega_{\nu\lambda\rho]} = 0$,
$\omega$ satisfies $\cL_\xi \omega = 0$.

Next, using the definition (2.1) and Equ. (2.8), it is straightforward
to show that
\beqa
\nabla_\mu \nabla_\nu V &=& - 2R_{\mu\lambda\nu\rho} \xi^\lambda \xi^\rho
+ 2V^{-2} [\sigma_{\mu\lambda} \sigma_\nu{}^\lambda 
- (\nabla^\lambda V) \xi_{(\mu} \sigma_{\nu)\lambda} \no \\
&& \mbox{} + \frac{1}{4} V \nabla_\mu V \nabla_\nu V + 
\frac{1}{4} \xi_\mu \xi_\nu (\nabla V)^2]
\eeqa
and whence
\beq
\nabla^\mu\nabla_\mu V = - 2R_{\mu\nu} \xi^\mu \xi^\nu
 + V^{-1}(\nabla V)^2 + 2V^{-2}\sigma_{\mu\nu} \sigma^{\mu\nu}.
\eeq
Now recall (see e.g. the Appendix of \cite{GeS}) that there is a $1-1$
correspondence between tensor fields on $M$ with vanishing Lie derivative
with respect to $\xi^\mu$ and such that all their contractions with
$\xi^\mu$ and $\xi_\mu$ are zero -- and ones of the same type on $N$.
In the case of covariant tensor fields on $N$, this correspondence is the
same as pull-back under $\pi$. Examples of such tensor fields on $M$ are
the scalar field $V$, the symmetric tensor field
\beq
h_{\mu\nu} := g_{\mu\nu} - V^{-1} \xi_\mu \xi_\nu 
\eeq
and the 2-form $\sigma_{\mu\nu} = \omega_{\mu\nu\lambda} \xi^\lambda$.
Note that $\sigma_{\mu\nu}$ can also be written as
\beq
\sigma_{\mu\nu} = V h_\mu{}^{\mu'} h_\nu{}^{\nu'} \nabla_{\mu'}
\xi_{\nu'} .
\eeq
The tensor $h_{\mu\nu}$ is, of course, the natural Riemannian metric on
$N$. The covariant derivative $D_\mu$ associated with $h_{\mu\nu}$ acting,
say, on a covector $X_\mu$ living on $N$, is given by
\beq
D_\mu X_\nu = h_\mu{}^{\mu'} h_\nu{}^{\nu'} \nabla_{\mu'} X_{\nu'}.
\eeq
Denoting by $\R_{\mu\nu\lambda\sigma}$ the curvature associated with
$D_\mu$, we find, using Equ. (2.9), that
\beq
\R_{\mu\nu\lambda\sigma} = h_\mu{}^{\mu'} h_\nu{}^{\nu'} 
h_\lambda{}^{\lambda'} h_\sigma{}^{\sigma'} R_{\mu'\nu'\lambda'\sigma'}
+ 2V^{-3} \sigma_{\mu\nu} \sigma_{\lambda\rho} 
- V^{-3} (\sigma_{\lambda[\mu} \sigma_{\nu ]\rho} -
\sigma_{\rho[\mu} \sigma_{\nu]\lambda}).
\eeq
Since $N$ is 3-dimensional, there holds
\beq
\sigma_{\mu[\nu} \sigma_{\lambda\rho]} = 0 ,
\eeq
so that
\beq
\R_{\mu\nu\lambda\rho} = h_\mu{}^{\mu'} h_\nu{}^{\nu'} 
h_\lambda{}^{\lambda'} h_\rho{}^{\rho'} 
R_{\mu'\nu'\lambda'\rho'} + 3V^{-3} \sigma_{\mu\nu} \sigma_{\lambda\rho}.
\eeq
Thus
\beq
\R_{\mu\nu} = h_\mu{}^{\mu'} h_\nu{}^{\nu'} R_{\mu'\nu'} -
V^{-1} R_{\mu\nu'\lambda\rho'} \xi^{\nu'} \xi^{\rho'} +
3 V^{-3} \sigma_{\mu\lambda} \sigma_\nu{}^\lambda .
\eeq
Using (2.18), Equ. (2.26) finally leads to
\beq
\R_{\mu\nu} = h_\mu{}^{\mu'} h_\nu{}^{\nu'} R_{\mu'\nu'} 
+ \frac{1}{2} V^{-1} D_\mu D_\nu V 
+ 2V^{-3} \sigma_{\mu\lambda} \sigma_\nu{}^\lambda 
- \frac{1}{4} V^{-2} (D_\mu V)(D_\nu V).
\eeq
From (2.19) we deduce that
\beq
D^2V := h^{\mu\nu} D_\mu D_\nu V = -2 R_{\mu\nu} \xi^\mu \xi^\nu +
\frac{1}{2} V^{-1} (DV)^2 + 2V^{-2} \sigma_{\mu\nu} \sigma^{\mu\nu}.
\eeq
We now make the following observation: when $\tau_{\mu \ldots \lambda}$
is an arbitrary tensor on $N$, there holds
\beq
h_\nu{}^{\nu'} \ldots h_\lambda{}^{\lambda'} \nabla^\mu
\tau_{\mu\nu' \ldots \lambda'} = (-V)^{-1/2} D^\mu [(-V)^{1/2}
\tau_{\mu\nu \ldots \lambda}].
\eeq
Applying (2.29) to (2.15) it follows that
\beq
(-V)^{-1/2} D^\mu [(-V)^{1/2} \sigma_{\mu\nu}] = h_\nu{}^{\nu'}
R_{\nu'\mu} \xi^\mu .
\eeq
Finally, projecting (2.10) down to $N$, it follows that
\beq
D_{[\mu} (V^{-2} \sigma_{\nu \rho]}) = 0 .
\eeq

Given the spacetime $(M,g_{\mu\nu})$ with the Killing vector $\xi^\mu$,
under the conditions stated at the beginning of this section, there are
coordinates $(t,x^i)$ on $M$, such that the canonical projection $\pi$
takes the form $\pi : (t,x^i) \mapsto (x^i)$, with $x^i$ local 
coordinates on $N$ and such that the Killing vector $\xi^\mu$ takes the
form $\xi = \partial/\partial t$. In terms of such coordinates tensor
fields on $N$, say $\tau_{i \ldots j}(x)$, can be viewed as the tensor
fields
\beq
\tau_{\mu \ldots \nu}(t,x) = \delta_\mu{}^i \ldots \delta_\nu{}^j
\tau_{i \ldots j}(x) .
\eeq
Since $\xi_\mu \xi^\mu = V$, there holds
\beq
\xi_\mu dx^\mu = V(dt + \vp_i dx^i),
\eeq
for some 1-form $\vp_i$. 
Note that, in the tangent space at each point $(t,x^i) \in M$, the
$g_{\mu\nu}$-orthogonal complement of $\xi_\mu$ is spanned by 
$\vp_i \partial/\partial t + \partial/\partial x^i$ and the orthogonal
complement of $\xi_\mu$ in the cotangent space is spanned by $dx^i$.
From the definition
$h_{\mu\nu} = g_{\mu\nu} - V^{-1} \xi_\mu \xi_\nu$ it follows that
\beq
g_{\mu\nu} dx^\mu dx^\nu = V(dt + \vp_i dx^i)^2 + h_{ij} dx^i dx^j ,
\eeq
where $V$, $\vp_i$, $h_{ij}$ on the r.h. side of (2.34) are all independent
of $t$. It is now straightforward to check that
\beq
\sigma_{\mu\nu} dx^\mu dx^\nu = 3(\xi_{[\mu} \nabla_\nu \xi_{\lambda]})
dx^\mu dx^\nu = V^2 \partial_{[i} \vp_{j]} dx^i dx^j .
\eeq
Thus $\sigma_{\mu\nu}$, viewed as a tensor on $N$, is given by
\beq
\sigma_{ij} = V^2 \partial_{[i} \vp_{j]} .
\eeq
In the static case $t$ can be chosen so that $\vp_i = 0$.

Conversely, let us start from the 3-manifold $(N,h_{ij},V,\sigma_{ij})$
with Riemannian metric $h_{ij}$, a negative scalar field $V$ and the 
2-form $\sigma_{ij}$, subject to
\beq
D_{[i} (V^{-2} \sigma_{jk]} ) = 0 ,
\eeq
which corresponds to (2.31). Suppose, moreover, that $N$ has trivial second 
cohomology. Then there exists a covector $\vp_i$ on $N$ with
\beq
\sigma_{ij} = V^2 D_{[i} \vp_{j]} .
\eeq
Define $M = \{t \in {\bf R}\} \times N$ and define on $N$ the Lorentz
metric $g_{\mu\nu}$ by Equ. (2.34) and $\xi^\mu$ by 
$\xi = \partial/\partial t$. Then one checks that $\xi_\mu \xi^\mu = V$,
that, under the projection $\pi : M \ra N$, $h_{\mu\nu}$ is the pull-back
of $h_{ij}$ and that $\sigma_{\mu\nu}$ is the pull-back of 
$\sigma_{ij} = V^2 D_{[i} \vp_{j]}$. The fact that the product structure
of $M$ as $M = {\bf R}^1 \times N$ is not natural is reflected in the
above construction by the fact that $\vp_i$, solving (2.38), is given
only up to $\vp_i \mapsto \bar \vp_i = \vp_i + D_i F$, with $F$ a scalar
field on $N$. Under this change $g_{\mu\nu}$ given Equ. (2.34) remains
unchanged only when we set $t \mapsto \bar t = t - F$.

Given the fields $(h_{ij}, V, \sigma_{ij})$ on $N$, we can define the
fields $r,r_i,r_{ij}$ by the following equations:
\beqa
D^2 V &=& -2r + \frac{1}{2} V^{-1}(DV)^2 + 2V^{-2} \sigma_{ij} \sigma^{ij} \\
D^i [(-V)^{-1/2} \sigma_{ij}] &= & (-V)^{1/2} r_j \\
\R_{ij} &=& r_{ij} + \frac{1}{2} V^{-1} D_i D_j V + 2V^{-3} \sigma_{ik}
\sigma_j{}^k - \frac{1}{4} V^{-2} (D_i V)(D_jV) . \no \\
\eeqa
It then follows from our previous considerations that the spacetime
$(M,g_{\mu\nu})$ satisfies
\beq
R_{\mu\nu} dx^\mu dx^\nu = r(dt + \vp_\ell dx^\ell)^2 + 2r_i dx^i
(dt + \vp_\ell dx^\ell) + r_{ij} dx^i dx^j .
\eeq
In particular, iff $r$, $r_i$, $r_{ij}$ are all zero, $(M,g_{\mu\nu})$ is 
a vacuum spacetime. In this case we refer to (2.39,40,41) as `the
vacuum equations'.

For later use we record another form of the field equations
\beq
G_{\mu\nu} = \kappa T_{\mu\nu},
\eeq
where
\beq
T_{\mu\nu} dx^\mu dx^\nu = \tau(dt + \vp_i dx^i)^2 + 2 \tau_i(dt + 
\vp_j dx^j)dx^i + \tau_{ij} dx^i dx^j,
\eeq
and where we set
\beq
g_{\mu\nu} dx^\mu dx^\nu = - e^{2U}(dt + \vp_i dx^i)^2 +
e^{-2U} \bar h_{ij} dx^i dx^j ,
\eeq
given by
\beqa
\bar D^2U &=& \frac{\kappa}{2} (e^{-4U} \tau + \bar \tau_\ell{}^\ell)
- e^{4U} \bar \omega_{ij} \bar \omega^{ij} \\
\bar D^i \bar \omega_{ij} &=& \kappa e^{-4U} \tau_j \\
\bar \R_{ij} &=& 2(D_iU)(D_jU) - 2e^{4U} \bar \omega_{ik} \bar \omega_j{}^k
+ \bar h_{ij} e^{4U} \bar \omega_{k\ell} \bar \omega^{k\ell} +
\kappa(\tau_{ij} - \bar h_{ij} \bar \tau_\ell{}^\ell). \no \\
\eeqa
Here
\beq
\bar \omega_{ij} = \omega_{ij} = \partial_{[i} \vp_{j]}
\eeq
and indices are raised with $\bar h^{ij}$.

\subsection{Axial Symmetry}

We now assume the existence of a second, spacelike Killing vector 
$\eta^\mu$ on $(M,g_{\mu\nu})$. There is the following identity
\beq
4 \nabla^\mu(\eta_{[\rho} \omega_{\mu\nu\lambda]}) = - \cL_\eta
\omega_{\nu\lambda\rho} + 6 \xi^\mu R_{\mu[\nu} \xi_\lambda \eta_{\rho]},
\eeq
where $\omega_{\mu\nu\lambda}$ is given by Equ. (2.2) and we have used (2.14).
Suppose, in addition, that $\xi$ and $\eta$ commute. Then the first term 
on the right in
(2.50) vanishes so that
\beq
4 \nabla^\mu(\eta_{[\rho} \omega_{\mu\nu\lambda]}) = 
6 \xi^\mu R_{\mu[\nu} \xi_\lambda \eta_{\rho]}.
\eeq
In an analogous manner
\beq
4 \nabla^\mu(\xi_{[\rho} \omega'_{\mu\nu\lambda]}) = 
- 6 \eta^\mu R_{\mu[\nu} \xi_\lambda \eta_{\rho]},
\eeq
where $\omega'_{\mu\nu\lambda}$ is given in terms of $\eta$ in the same
way as $\omega_{\mu\nu\lambda}$ is given in terms of $\xi$.
The r.h. sides of Equ.'s (2.51,52) are zero (at points where $\xi$ and $\eta$
are linearly independent) iff  the timelike 2-plane spanned by $\xi$ and $\eta$ is invariant under
$R_\mu{}^\nu$. These conditions will be satisfied when the energy momentum 
tensor is that of a rotating perfect fluid.
We now assume that $\eta^\mu$ has an axis, i.e. vanishes on a timelike
2-surface which is tangent to $\xi^\mu$. Then, and when the r.h. sides
of (2.51,52) are zero, it follows that
\beq
\eta_{[\rho} \omega_{\mu\nu\lambda]} = \xi_{[\rho} \omega'_{\mu\nu\lambda]}
= 0.
\eeq
The relations (2.53), in turn, are nothing but the conditions for the 2-plane
elements orthogonal to $\xi$ and $\eta$ to be integrable (``surface
transitivity of $\xi$ and $\eta$''). The above result is due to
Kundt and Tr\"umper \cite{KuTr}.

For the purposes of Sect. 4.2 we need to transcribe the relations
satisfied by $\eta^\mu$ on the quotient manifold $N$. Writing the 1-form
$\eta_\mu = g_{\mu\nu} \eta^\nu$ as
\beq
\eta_\mu dx^\mu = \eta(dt + \vp_i dx^i) + \eta_i dx^i,
\eeq
so that
\beq
\eta^\mu \frac{\partial}{\partial x^\mu} = (V^{-1} \eta - \vp_i \eta^i)
\frac{\partial}{\partial t} + \eta^i \frac{\partial}{\partial x^i},
\eeq
the Killing equations
\beq
\eta^\lambda \partial_\lambda g_{\mu\nu} + 2 g_{\lambda(\mu} \partial_{\nu)}
\eta^\lambda = 0
\eeq
are equivalent to
\beqa
\eta^i D_i V &=& 0 \\
2 \omega_{ij} \eta^j &=& D_i(V^{-1} \eta) \\
\cL_\eta h_{ij} &=& 0,
\eeqa
where $\omega_{ij} := D_{[i} \vp_{j]}$. The surface transitivity conditions
(2.53) get translated into
\beqa
\eta_{[i} D_j \eta_{k]} &=& 0 \\
\eta_{[i} \omega_{jk]} &=& 0 .
\eeqa
In particular, $\eta^i$ is a hypersurface-orthogonal Killing vector on
$(N,h_{ij})$.
Suppose, now, that the energy momentum tensor is that of a rigidly
rotating perfect fluid, i.e.
\beq
T_{\mu\nu} = (\rho + p) u_\mu u_\nu + p g_{\mu\nu} ,
\eeq
with
\beq
u_\mu = f(\xi_\mu + \Omega \eta_\mu), \qquad \Omega = \mbox{const,}
\eeq
and $f$ is chosen so that $u^\mu$ is future-pointing and
$u_\mu u^\mu = -1$. With this specialization the quantities
$\tau,\tau_i,\tau_{ij}$ entering in the field equation (2.46,47,48) become
\beqa
\tau &=& f^2(\rho + p)(-e^{2U} + \Omega \eta)^2 - p e^{2U} \\
\tau_i &=& f^2(- e^{2U} + \Omega \eta)(\rho + p) \Omega \eta_i \\
\tau_{ij} &=& p e^{-2U} \bar h_{ij} + f^2 \Omega^2(\rho + p) \eta_i \eta_j,
\eeqa
where $\eta_i = h_{ij} \eta^j$. The normalization factor $f$ is given by
\beq
f = [e^{-2U} (-e^{2U} + \Omega \eta)^2 - \Omega^2 \eta_\ell 
\eta^\ell]^{-1/2}.
\eeq
The field equations have to be supplemented by the Killing relations
(2.57,58,59). Note that these imply that $\rho$ and $p$ are invariant under
$\eta^i$ (in addition of course to being invariant under $\partial/\partial t$).
Under these circumstances the contracted Bianchi identities, which imply
that
\beq
\nabla_\mu T^{\mu\nu} = 0,
\eeq
boil down to the relation
\beq
(\rho + p) f^{-1} D_i f = D_i p,
\eeq
the remaining condition, namely $\bar D^j(e^{-4U} \tau_j) = 0$, being
identically satisfied.

\subsection{Asymptotic Flatness -- Lichnerowicz Theorems}

Before stating the conditions for stationary spacetimes to be asymptotically
flat, we elaborate somewhat more on the vacuum field equations. First
recall from (2.17) that, when $M$ (or equivalently: $N$) is simply connected,
there exists a field $\omega$ on $N$ such that
\beq
\sigma_{ij} = \frac{1}{2} (-V)^{1/2} \ve_{ij}{}^k D_k \omega ,
\eeq
where we have used (2.16) and
\beq
\ve_{ijk} dx^i dx^j dx^k = (-V)^{1/2} \xi^\mu \ve_{\mu\nu\lambda\sigma}
dx^\nu dx^\lambda dx^\sigma .
\eeq
(Of course, the existence of $\omega$ could have also been inferred from
(2.40) for $r_i = 0$.) We now rewrite the vacuum equations in terms of the 
conformally rescaled metric $\bar h_{ij}$ (see (2.45)), given by
\beq
\bar h_{ij} = (-V) h_{ij} .
\eeq
Then (2.39,40,41), together with (2.70) lead to
\beqa
\bar D^2 V &=& V^{-1} (\bar D V)^2 - V^{-1} (\bar D \omega)^2 \\
\bar D^2 \omega &=& 2V^{-1} (\bar D \omega)(\bar D V)
\eeqa
and
\beq
\bar \R_{ij} = \frac{1}{2} V^{-2} [(D_iV)(D_j V) +
(D_i \omega)(D_j \omega)] ,
\eeq
or
\beq
\bar G_{ij} = \frac{1}{2} V^{-2} \left\{ (D_i V)(D_jV) +
(D_i \omega)(D_j \omega) - \frac{1}{2} \bar h_{ij}
[(\bar D V)^2 + (\bar D \omega)^2]\right\} .
\eeq
We can now give an interesting geometric interpretation of the vacuum
equations (2.73,74,76). Namely, let $\cP$ be the Poincar\'e half-plane
with metric $q_{AB}$ given by
\beq
q_{AB} dz^A dz^B = V^{-2}(dV^2 + d\omega^2) \quad
(V > 0, -\infty < \omega < \infty).
\eeq
Viewing $(z^1(x), z^2(x)) = (V(x),\omega(x))$ as a map from
$(N,\bar h_{ij})$ to $(\cP,q_{AB})$, one easily checks that Equ.'s (2.73,74)
are exactly the conditions in order for this map to be harmonic,
in other
words
\beq
\bar D^2 z^A + \Gamma^A_{BC} z^B{}_{,j} z^C{}_{,j} \bar h^{ij} = 0 ,
\eeq
where $\Gamma^A_{BC}$ denotes the Christoffel symbols of $q_{AB}$,
composed with $z^C(x)$. The metric $\bar h_{ij}(x)$, of course, is not
given, but has to satisfy (2.51). The r.h. side of Equ. (2.51), in turn,
is nothing but the energy momentum tensor of the harmonic map.
$(\cP,q_{AB})$ can also be viewed as a spacelike hyperboloid in
(2+1)-dimensional Minkowski space. Namely, define fields
\beq
\Phi_M = \frac{V^2 + \omega^2 -1}{-4V}, \qquad
\Phi_S = - \frac{\omega}{2V}, \qquad
\Phi_K = - \frac{V^2 + \omega^2 +1}{4V}.
\eeq
Then
\beq
- \Phi^2_K + \Phi^2_M + \Phi^2_S = - \frac{1}{4} .
\eeq
Viewing $(\Phi_K,\Phi_M,\Phi_S)$ as coordinates on ${\bf R}^3$ with
Lorentz metric $4(-d \Phi^2_K + d\Phi^2_M + d\Phi^2_S)$, the induced
metric under the map (2.79) is nothing but $q_{AB}$. The fields
$\Phi_M,\Phi_S$ are the potentials first introduced by Hansen \cite{Hansen} 
which we shall use in Sect.~3.1 and 3.2.

As with any harmonic map, we can associate a conserved current on $N$ with
any Killing vector on the target space $\cP$. Since $\cP$ has $SO(2,1)$ as
isometry group, there are three independent such Killing vectors, namely
\beqa
\st{1}{\eta}{}^A \;\frac{\partial}{\partial z^A} &=&
\frac{\partial}{\partial \omega} \\
\st{2}{\eta}{}^A \;\frac{\partial}{\partial z^A} &=&
V \frac{\partial}{\partial V} + \omega \frac{\partial}{\partial \omega} \\
\st{3}{\eta}{}^A \;\frac{\partial}{\partial z^A} &=&
\omega V \frac{\partial}{\partial V} + \frac{1}{2}
(\omega^2 - V^2)  \frac{\partial}{\partial \omega} .
\eeqa
We note in passing that the $SO(2,1)$ isometry of $\cP$
is closely related to the ``Ehlers transformation'' discussed in the
 article by Maison in this volume.
The conserved current $j_i$ associated with any Killing vector $\eta^A$ on
$\cP$ is given by
\beq
j_i = z^A{}_{,i} \eta^B q_{AB} .
\eeq
Hence
\beqa
\st{1}{j_i} &=& V^{-2} D_i \omega \\
\st{2}{j_i} &=& V^{-1} D_i V + V^{-2} \omega D_i \omega \\
\st{3}{j_i} &=& V^{-1} \omega D_i V + (2V)^{-2} 
(\omega^2 - V^2) D_i \omega 
\eeqa
are all divergence-free on $(N,\bar h_{ij})$.
By (2.70) and (2.38), $\st{1}{j_i}$ is also equal to
\beq
\st{1}{j_i} = \bar \ve_i{}^{jk} D_j \vp_k ,
\eeq
and so the ``charge'' associated with $\st{1}{j_i}$ is always zero. In
the asymptotically flat we shall turn to later, (2.87) will be identically zero.

The quantity (2.86) has the following spacetime interpretation (compare
\cite{GeS}). Let $\Sigma$
be a 2-surface in $M$ which projects down to a smooth 2-surface on $N$.
Then there exist local coordinates $(x^\mu) = (t,x^i)$ such that $\Sigma$
is given by
\beq
x^\mu(y^A) = (0,x^i(y^A)), \qquad A = 1,2.
\eeq
Now integrate the quantity $\ve_{\mu\nu\rho\sigma} \nabla^\rho \xi^\sigma$
over $\Sigma$. After some computation one finds
\beq
\ve_{\mu\nu\rho\sigma} (\nabla^\rho \xi^\sigma) 
\frac{\partial x^\mu}{\partial y^1} \frac{\partial x^\nu}{\partial y^2}
= (-V)^{-1/2} (\partial_i V - 2 \sigma_{ij} \vp^j) \ve^i{}_{k\ell}
\frac{\partial x^k}{\partial y^1} \frac{\partial x^\ell}{\partial y^2}
\eeq
where, as before, $\sigma_{ij} = V^2 \partial_{[i} \vp_{j]}$. Now, using
(2.70),
\beq
-2(-V)^{-1/2} \sigma_{ij} \vp^j + (-V)^{-3/2} \omega D_i \omega =
D^j(\omega \ve_{ij}{}^k \vp_k).
\eeq
Thus, when $\Sigma$ is closed, the integral $I$ of the expression (2.91) is
given by $(\bar h_{ij} = (-V) h_{ij})$
\beq
I = \int_\Sigma (-V)^{-1}(D_i V + V^{-1} \omega D_i \omega)d \bar S^i.
\eeq
The fact that this integral in vacuum only depends on the homology class
of $\Sigma$ arises, in the spacetime picture, from the fact that
\beq
\nabla^\mu \nabla_{[\mu} \xi_{\nu]} = 0, \mbox{ when } R_{\mu\nu} = 0.
\eeq
The quantity
\beq
M = \frac{1}{8\pi} \; I
\eeq
is called the Komar mass of $(M,g_{\mu\nu})$. For the Schwarzschild 
solution it coincides with the Schwarzschild mass when the ``outward''
orientation is chosen for $d \bar S^i$.

We now come to the
\subsection*{Boundary Conditions}
Recall that we require $(M,g_{\mu\nu})$ to be connected, simply connected
and chronological. Let, in addition, $M$ contain a compact subset $K$ and
let $M \setminus K$ be an ``asymptotically flat end''. (The results of this
subsection will remain to be true if $M \setminus K$ consists of finitely
many asymptotic ends.) This means that $M \setminus K$ should be
diffeomorphic to $M_R$ $(R > 0)$ with
\beq
M_R = \{(x^0,x^i) \in {\bf R}^1 \times ({\bf R}^3 \setminus B(R))\}
\eeq
with $B(R)$ a closed ball of radius $R$. In terms of this diffeomorphism,
the metric $g_{\mu\nu}$ in $M \setminus K$ has to satisfy that there
exists a constant $C > 0$ such that (see \cite{BeChr1})
\beq
|g_{\mu\nu}| + |g^{\mu\nu}| + r^\alpha |g_{\mu\nu} - \eta_{\mu\nu}| +
r^{1+\alpha} |\partial_\sigma g_{\mu\nu}| +
r^{2+\alpha} |\partial_\sigma \partial_\rho g_{\mu\nu}| \leq C
\eeq
\beq
g_{00} \leq - C^{-1}, \qquad g^{00} \leq - C^{-1} 
\eeq
\beq
\forall \; X^i \in {\bf R}^3 \quad g_{ij} X^i X^j \geq C^{-1}
\sum (X^i)^2.
\eeq
We assume $\alpha > 1/2$. Furthermore we require $R_{\mu\nu}$ to be zero
in $M \setminus K$. (This latter condition could be considerably relaxed.)
It now follows that the level set $x^0 = 0$ is a spacelike submanifold
of $M \setminus K$ which has a finite ADM-momentum $p^\mu$ (see \cite{BeChr1}).
If $p^\mu$ is a timelike vector (which it will be for `reasonable' matter
except in the vacuum case), it now follows from the timelike character
of $\xi^\mu$ that it has to be an asymptotic time translation, i.e.
\beq
|\xi^\mu - A^\mu| + r |\partial_\sigma \xi^\mu| + r^2 |\partial_\rho 
\partial_\sigma \xi^\mu| \leq C r^{-\alpha} ,
\eeq
where the constants $A^\mu$ satisfy
\beq
A^\mu A^\nu \eta_{\mu\nu} < 0
\eeq
(see \cite{BeChr1}). Furthermore it follows from \cite{BeChr1}, that in $M \setminus K$, or 
a subset thereof diffeomorphic to $M_{R'}$ for sufficiently large
$R' > R$, there are coordinates $(t,y^i)$ in terms of which $g_{\mu\nu}$ 
is again asymptotically flat with the same $\alpha > 1/2$ and so that
$\xi^\mu$ is of the form $\xi^\mu \; \partial/\partial x^\mu =
\partial/\partial t$. Hence, in the coordinates $(t,y^i)$, which we now 
call $(t,x^i)$, the metric
\beq
g_{\mu\nu} dx^\mu dx^\nu = V(dt + \vp_i dx^i)^2 + h_{ij} dx^i dx^j
\eeq
satisfies
\beqa
|V+1| + r|\partial_i V| + r^2|\partial_i \partial_j V| 
&\leq& C r^{-\alpha} \\
|\vp_i| + r|\partial_j \vp_i| + r^2|\partial_k \partial_j \vp_i|
&\leq& C r^{-\alpha} \\
|h_{ij} - \delta_{ij}| + r|\partial_k h_{ij}| + r^2|\partial_k 
\partial_\ell h_{ij}| &\leq& C r^{-\alpha}
\eeqa
in $M \setminus \C$. It follows that
\beq
r |\sigma_{ij}| + r^2|\partial_k \sigma_{ij}| \leq C r^{-\alpha}.
\eeq
We remark that the time coordinate $t$, which is at first only defined 
in the open subset $M \setminus K$ of $M$, can be (Kobayashi--Nomizu 
\cite{KoNo})
extended to a smooth global cross section of $\pi : M \ra N$.

We now state and prove two uniqueness theorems due to Lichnerowicz \cite{Li},
which are basic for the theory of stationary solutions.

\paragraph{Staticity theorem:} Let $(M,g_{\mu\nu},\xi^\lambda)$ be
asymptotically flat with $\alpha > 1/2$, $\xi^\mu$ be an asymptotic time
translation. If the matter is non-rotating relative to $\xi^\mu$, i.e.
$r_i$ in Equ. (2.40) is zero, then the spacetime is static.

 \paragraph{Proof:} From (2.40) we have
\beq
D^i [(-V)^{-1/2} \sigma_{ij}] = 0 .
\eeq
Contract Equ. (2.106) with $\vp^j$, using $\sigma_{ij} = V^2 D_{[i}\vp_{j]}$.
It follows that
\beq
D^i[(-V)^{-1/2} \sigma_{ij} \vp^j] = (-V)^{-5/2} \sigma_{ij} \sigma^{ij}.
\eeq
Now integrate Equ. (2.107) over $N$. Since the term in brackets on the left
is $O(r^{-2-2\alpha})$, the boundary term at infinity gives zero.
Consequently $\sigma_{ij} = 0 \Ra \omega_{\mu\nu\lambda} = 0$.
\paragraph{Remark:} Since $\sigma_{ij} = 0$, the field $\vp_i$ is of the
form $\vp_i = D_i F$, where $F = O(r^{1-\alpha})$. In the coordinates
$\bar t = t - F$, $g_{\mu\nu}$ takes the form
\beq
g_{\mu\nu} dx^\mu dx^\nu = V dt^2 + h_{ij} dx^i dx^j.
\eeq
\paragraph{Vacuum theorem:} Let $(M,g_{\mu\nu},\xi^\lambda)$ satisfy the
conditions in the staticity theorem and let $(M,g_{\mu\nu})$ in addition
be vacuum. Then $(M,g_{\mu\nu})$ is the Minkowski space.
\paragraph{Proof:} Firstly, by the staticity theorem, we have that
$\sigma_{ij} = 0$. Using this in Equ. (2.39) for $r = 0$ we have with
$v := (-V)^{1/2}$ that
\beq
D^2 v = 0 .
\eeq
By the maximum principle, or multiplying (2.109) by $\mu$, integrating
by parts and using $\mu - 1 = O(r^{-\alpha})$,
$\partial_i \mu = O(r^{-1-\alpha})$, we infer that $\mu \equiv 1$.
Now Equ.~(2.41) implies $\R_{ij} = 0 \Ra \R_{ijk\ell} = 0$ since
$\dim N = 3$. Since $N$ is simply connected, it follows that
$(N,h_{ij})$ is flat ${\bf R}^3$. Thus
\beq
g_{\mu\nu} dx^\mu dx^\nu = - dt^2 + \delta_{ij} dx^i dx^j .
\eeq

\subsection{Newtonian Limit}
 Ehlers showed (unpublished, see \cite{Lott})  that one can write the field equation
containing a parameter $\lambda=c^{-2}$ such that the equation remain
meaningful for  $\lambda=0$ and then they are equivalent to the
Newtonian equations. The variables for which this is true in the time dependent
case have to be chosen in a quite sophisticated way. The stationary case can be
treated in a direct and simple  way as follows. 

We  write the metric as 
\begin{equation}
g_{\mu\nu}dx^\mu dx^\nu =-e^{-\frac{2U}{c^2}}(cdt +\vp_idx^i)^2+
e^{\frac{2U}{c^2}}\bar h_{ik}dx^idx^k
\end{equation}
where we inserted $``c"$  by dimensional analysis. The field equations
decomposed  in section 2.1
\begin{equation}
R_{\mu\nu}=\frac{8\pi G}{c^4}(T_{\mu\nu}-\frac{1}{2}Tg_{\mu\nu})
\end{equation}
for the energy  momentum tensor 
\begin{equation}
T_{\mu\nu}=c^2\tau(cdt +\vp_idx^i)^2 + 2 c \tau_i(cdt +\vp_jdx^j)dx^i
+\tau_{ij}dx^idx^j
\end{equation}
become
\begin{equation}
\bar D^2U=4\pi
G(e^{-\frac{4U}{c^2}}\tau+c^{-2}\bar\tau_l^l)-e^{\frac{4U}{c^2}}
\bar\omega_{ij}\bar\omega^{ij}
\end{equation}
\begin{equation}
\bar D^i\bar\omega_{ij}=8\pi G
c^{-3}e^{-\frac{4U}{c^2}}\tau_j
\end{equation}
\begin{equation}
\bar R_{ij}=2c^{-4}D_iUD_jU 
-2e^{\frac{4U}{c^2}}\bar\omega_{ik}\bar\omega_j{}^k+
\bar h_{ij}e^{\frac{4U}{c^2}}\bar\omega_{kl}\bar\omega^{kl}
+8\pi Gc^{-4}(\tau_{ij}-\bar h _{ij} \bar\tau_l^l)
\end{equation}
Considered as equations for $U, \bar h_{ij}, \phi_i, \tau, \tau_i,
\tau_{ij}$, the equations (2.114--116) have a limit for $c\to\infty.$ 

In the static case the limit is
\begin{equation}
\bar D^2 U=4\pi G \tau
\end{equation}
\begin{equation} 
\bar R_{ik}=0
\end{equation}
Hence we obtain immediately that the metric $\bar h_{ik}$ of the quotient is flat
and therefore (2.117) is the Poisson equation of Newton's theory. The connection
has also a limit and the only non vanishing Christoffel symbol is
$\Gamma^i_{tt}= \bar D^iU$. The equation of motion $\nabla_\nu
T^{\mu\nu}=0$
becomes in the limit the Newtonian equilibrium condition
$\bar D^j\tau_{ij}=-\tau D_j U$.

Now to the stationary case:
Because the right hand side of (2.115) vanishes in  the Newtonian limit,
the Lichnerowicz theorem implies that
$\omega_{ik}=0$
 which in turn implies by (2.116) that $\bar R_{ik}=0$ whence the metric on 
the quotient is again flat.

For the metric written in the form (1) the connection has no limit for
$c\to\infty$. If we use however as a consequence of the field equations
that
$\omega_{ij}=0$, the connection has a limit and the equations of motion
become
the Newtonian equilibrium conditions, i.e.
\begin{equation}
\bar D^i\tau_i=0\ \ \ , \ \ \ \bar D^j\tau_{ij}=-\tau D_j U\ . 
\end{equation}
\subsection{Existence Issues and the Newtonian Limit}
The fact that the equations can be written to contain $\lambda=c^{-2}$ in such
a way that they are analytic in $\lambda$ and are the Newtonian equations for
$\lambda=0$, suggests to use this structure for existence theory. In this section
we will make some remarks about the static case. In section 4.2 an existence
theorem for a rigidly rotating body by Heilig will be discussed which exploits
the fact that the equations have a nice Newtonian limit. 

To obtain partial differential equations for which there is an existence theory
we write (2.104) and (2.105) in the static case in harmonic coordinates on $N$, defined
by  $\bar D^2x^i=0$, for the unknowns $U$ and $Z^{ij}$ defined by $\bar
h^{ij} =\delta^{ij} + \lambda^2 Z^{ij}$ and obtain:
\beqa
\Delta U:=\delta^{ij}\partial_i\partial_jU &=& 
4\pi G \tau +A(\lambda,\tau,\tau_{ij},Z^{ij}) \\
\Delta Z^{ij} &=& -4\partial_kU \partial_lU\delta^{ik}\delta^{lj} 
-16\pi G (\bar\tau^{ij}-\bar\tau\delta^{ij}) \no \\
&& \mbox{} + \lambda^2B^{ij}(\lambda,\tau,\tau_{kl},Z^{kl},\partial_m Z^{kl},
\partial_m \partial_n
Z^{kl})
\eeqa
here we used the well known expression for the Ricci tensor in harmonic
coordinates
\begin{equation}
\bar R^{ij}=-\frac{1}{2}\bar h^{kl}\partial_k\partial_l\bar h^{ij}
+ H^{ij}(\partial \bar h,\partial\bar h)
\end{equation}
where $H^{ij}$ is quadratic in the first derivatives of $\bar h^{ij}$.
As usual we call (2.120), (2.121) the reduced field equations. These form a
quasilinear elliptic system with the property that for given  small sources
$\tau,\tau_{ij}$ of compact support and small $\lambda$ there exist unique
solutions $U, Z^{ij}$ which tend to $0$ at infinity.

In particular we can choose for $\tau,\tau_{ij}$ a Newtonian solution and
determine then for small $\lambda$ a relativistic solution of the reduced field
equations which have a Newtonian limit. Is is to be expected that the solution
will be analytic in $\lambda$. Then the Taylor expansion in $\lambda$ can be
considered as a converging post Newtonian expansion.

A solution of the reduced field equations is only solution of the field equation
if it satisfies the harmonicity condition or equivalently if $\nabla_\mu
T^{\mu\nu}=0$ holds. To solve the reduced equations and the equation of motion
is a much harder problem. It makes only sense once a matter model is chosen. 

In the static case only matter models of elasticity lead to new interesting
problems because, as we will see in sections 5.2 , static fluids are spherically
symmetric and can be investigated by ordinary differential equations (see
Sect.~5.3). Some remarks on static, small self gravitating bodies can be found in
Sect.~5.1.

For a stationary rigidly rotating fluid Heilig has given an existence theorem
by perturbing away from a Newtonian solution. We will describe this result in
Sect.~4.2.   

\section{Far Fields}

\subsection{Far-Field Expansions}
While, as we have seen, little is known so far about globally regular,
asymptotically flat solutions to the stationary field equations with
reasonable matter sources, there is an almost complete understanding of
the behaviour of general asymptotically flat solutions near spatial
infinity, which we now describe.

Here the quotient manifold $N$ is of the form
$$
N = {\bf R}^3 \setminus B(R).
$$
On $N$ there are given the fields $(h_{ij},V,\vp_i)$ satisfying (2.39,40,41).
The whole discussion is ``local-at-infinity''. In particular one has to
allow for $R$ in $B(R)$ to be made suitably large, as one proceeds. We will
do so tacitly without changing the letter ``$R$''. 
The Einstein equations are given by
\beqa
\bar D^2 V &=& V^{-1} (\bar DV)^2 - V^{-1}(\bar D\omega)^2 \\
\bar D^2 \omega &=& 2V^{-1}(\bar D \omega)(\bar DV) \\
\bar G_{ij} &=& \frac{1}{2} V^{-2} \{(D_iV)(D_jV) + (D_i\omega)(D_j\omega) -
\frac{1}{2} \bar h_{ij} [(\bar DV)^2 + (\bar D\omega)^2]\}.
\eeqa
By the asymptotic conditions (2.102--104), $\omega$ tends to a constant at infinity.
Subtracting this from $\omega$, and calling the result again $\omega$, we
find that
\beq
|\omega| + r |\partial_i \omega| + r^2 |\partial_i \partial_j \omega|
\leq C r^{-\alpha}.
\eeq
In short, we have that
\beq
V = - 1 + O(r^{-\alpha}), \qquad
\omega = O(r^{-\alpha}), \qquad
\bar h_{ij} - \delta_{ij} = O(r^{-\alpha}), \qquad 1 > \alpha > 1/2
\eeq
and that these relations may be differentiated twice. The condition
$\alpha > 1/2$ could be relaxed (see Kennefick and \'O Murchadha 
\cite{KeO}). It now follows
that
\beqa
\Delta V &=& O(r^{-2-2\alpha}) \\
\Delta \omega &=& O(r^{-2-2\alpha}).
\eeqa
Since the r.h. sides of (3.6,7) decay stronger than $O(r^{-3})$, it
follows from standard results in potential theory \cite{Kellogg,SiB} 
that there exist constants $M,S$ such that
\beqa
V &=& - 1 + \frac{2M}{r} + O(r^{-1-\alpha}) \\
\omega &=& \frac{2S}{r} + O(r^{-1-\alpha}).
\eeqa
But the existence of $\vp_i$ in Equ. (2.36) implies that $S$ has to be zero.

The equation (3.3), which involves second derivatives of the metric
$\bar h_{ij}$, yields
\beq
\Delta(k_{ij} - \frac{1}{2} \delta_{ij} k) - 2 \partial_{(i} \Gamma_{j)}
+ \delta_{ij} \partial_\ell \Gamma_\ell = O(r^{-2-2\alpha}),
\eeq
where $k_{ij} = \bar h_{ij} - \delta_{ij}$, $k = k_{ii}$ and
\beq
\Gamma_i = \partial_j k_{ij} - \frac{1}{2} \partial_i k.
\eeq
This equation can be viewed in two ways, both of which recur in the
higher-order steps leading to the theorem below. Firstly, in the gauge
where $\Gamma_i = 0$, i.e. the linearized harmonic gauge for $\bar h_{ij}$,
it is an elliptic equation, namely essentially the componentwise Laplace
equation, for the leading-order part of $k_{ij}$. Secondly, Equ. (3.10)
can be rewritten as
\beq
\ve_{i\ell m} \ve_{jnp} \partial_\ell \partial_n k_{mp} =
O(r^{-2-2\alpha}) ,
\eeq
which expresses the fact that the linearized Riemann tensor of $\bar h_{ij}$
decays faster than $O(r^{-3})$. Note that (3.12) makes essential use
of the three-dimensionality of space. It follows \cite{SiB} that there exists
$g_i = O(r^{1-\alpha})$ such that
\beq
k_{ij} = \partial_i g_j + \partial_j g_i + O(r^{-2\alpha}).
\eeq
Thus the leading-order contribution to the metric $\bar h_{ij}$ is 
``pure gauge''.

To next order in $1/r$ one finds that there is a gauge, namely
\beq
\Gamma_i = O(r^{-3-\alpha}),
\eeq
for which
\beqa
V &=& - 1 + \frac{2M}{r} - \frac{2M_ix^i}{r^3} + \frac{2M^2}{r^2} +
O(r^{-1-2\alpha}) \\
\omega &=& \frac{2S_ix^i}{r^3} + O(r^{-1-2\alpha})
\eeqa
and for which $\bar h_{ij}$ can be brought into the form
\beq
k_{ij} = - \frac{M^2(\delta_{ij}r^2 - x_ix_j)}{r^4} + O(r^{-1-2\alpha}).
\eeq
In the above $M, M_i, S_i$ are constants. All indices are lowered and
raised with $\delta_{ij}$. When $M \neq 0$ one can, by a rigid translation,
arrange for $M_i = 0$. In that case the metric $g_{\mu\nu}$ obtained
from (3.15--17) coincides, to order $1/r^2$, with that of the Kerr
spacetime with $|S| = - Ma$, $M$ being the mass and $a$ being the Kerr
parameter.

In order to extend the above result to higher orders in $1/r$, it is
convenient to replace $(V,\omega)$ by some other choice of scalar
potentials. One choice, due to Hansen \cite{Hansen}, is to set (see
Equ. (2.79)
\beqa
\phi_M &=& - \frac{V^2 + \omega^2 - 1}{4V} \\
\phi_S &=& - \frac{\omega}{2V} \\
\phi_K &=& - \frac{V^2 + \omega^2 +1}{4V}
\eeqa
It then turns out that $(\phi_\alpha) = (\phi_M,\phi_S,\phi_K)$ all
satisfy
\beq
\bar D^2 \phi_\alpha = 2 \bar\R \phi_\alpha.
\eeq
Then one has \cite{SiB} the following
\paragraph{Theorem:} There exists a gauge, namely that where $\Gamma_i = 
O(r^{-m-1-\alpha})$, for which there are constants
$A \ldots,B \ldots, \ldots,G \ldots$ such that
\beqa
\phi_M &=& \sum_{\ell = 0}^{m-1}
\frac{E_{i_1 \ldots i_\ell} x^{i_1} \ldots x^{i_\ell}}{\ell! r^{2\ell+1}}
+ O^\infty(r^{-m+1-2\alpha}) \\
\phi_S &=& \sum_{\ell = 0}^{m-1}
\frac{F_{i_1 \ldots i_\ell} x^{i_1} \ldots x^{i_\ell}}{\ell! r^{2\ell+1}}
+ O^\infty(r^{-m+1-2\alpha}) \\
\phi_K &=& \frac{1}{2} + \sum_{\ell = 0}^{m-1}
\frac{G_{i_1 \ldots i_{\ell -1}} x^{i_1} \ldots x^{i_{\ell-1}}}{\ell! r^{2\ell}}
+ O^\infty(r^{-m+1-2\alpha}) 
\eeqa
Note that $E = M$,
\beqa
\bar h_{ij} &=& \delta_{ij} + \sum_{\ell =2}^m \left( 
\frac{x_ix_j A_{i_1 \ldots i_{\ell-2}} x^{i_1} \ldots x^{i_{\ell-2}}}
{r^{2\ell}} +
\frac{\delta_{ij} B_{i_1 \ldots i_{\ell-2}} x^{i_1} \ldots x^{i_{\ell-2}}}
{r^{2\ell-2}} \right. \no \\
&& \mbox{} + \left. 
\frac{x_{(i} C_{j)i_1 \ldots i_{\ell-3}} x^{i_1} \ldots x^{i_{\ell-3}}}
{r^{2\ell-2}} +
\frac{D_{iji_1 \ldots i_{\ell-4}} x^{i_1} \ldots x^{i_{\ell-4}}}
{r^{2\ell-4}} + 
O^\infty(r^{-m+1-2\alpha})\right). \no \\
\eeqa
All constants are symmetric in their $i_1 \ldots$ indices.
The constants $D$ are also symmetric in $i$ and $j$.
The constants $C\ldots$ appear only for $m \geq 3$, the constants $D\ldots$
only for $m \geq 4$.
The symbol $O^\infty(r^k)$ means that the quantity in question is of
$O(r^k)$, its derivative is $O(r^{k-1})$, a.s.o.
Furthermore all constants are determined by the tracefree parts of
$E \ldots,F \ldots$ in a way which does not depend on the solution at
hand. The tracefree parts of $E \ldots$ are the analogues of the Newtonian
multipole moments. The constants $F \ldots$ play an analogous role for
the ``angular-momentum aspect'', which does not have a Newtonian
counterpart. The three-metric $\bar h_{ij}$, for reasons explained after
(3.14), has no independent degrees of freedom.

This theorem shows, in essence, that any stationary, asymptotically flat
solution to the Einstein vacuum equations is uniquely determined by the
``moments'' $E \ldots,F \ldots$. However no statement concerning 
convergence of series like the ones appearing in (3.22--25) can be
made. In order to do that it is necessary to use ``conformal
compactification'' of three-space $N$.

\subsection{Conformal Treatment of Infinity, Multipole Moments}
Before turning to the situation G.R., it is instructive to recall the
Newtonian situation. Suppose we are given a Newtonian potential near
infinity, i.e. a function $\phi$ with
\beq
\Delta \phi = 0 \qquad \mbox{on } {\bf R}^3 \setminus B(R).
\eeq
Extending $\phi$ smoothly to all of ${\bf R}^3$, we thus have that
\beq
\Delta \phi = 4 \pi \rho \qquad \mbox{with } 
\rho \in C_0^\infty({\bf R}^3)
\eeq
and $\phi \ra 0$ at infinity. Thus $\phi$ is of the form
\beq
\phi(x) = - \int_{{\bf R}^3} \frac{\rho(x')}{|x-x'|} dx'.
\eeq
In ${\bf R}^3 \setminus B(R)$ this can (see e.g. \cite{Kellogg}) be expanded in a
standard fashion in powers of $1/r$. One obtains an expansion of the
form
\beq
\phi = \sum_{\ell = 0}^\infty 
\frac{E_{i_1 \ldots i_\ell} x^{i_1} \ldots x^{i_\ell}}{\ell! r^{2\ell+1}},
\eeq
with $E_{i_1 \ldots i_\ell}$ totally symmetric and tracefree. One shows
\cite{Kellogg,SiB} that this series converges absolutely and uniformly in
${\bf R}^3 \setminus B(R)$ for sufficiently large $R$.

As a warm-up for G.R. it is useful to rephrase the Newtonian situation
using ``conformal compactification''. First observe that there is a
positive smooth function $\Omega$ on $N = {\bf R}^3 \setminus B(R)$ with
the following properties. The metric
\beq
\wt h_{ij} = \Omega^2 \delta_{ij}
\eeq
extends to a smooth metric on the one-point compactification
\beq
\wt N = N \cup \{r = \infty\} = N \cup \{ \Lambda\},
\eeq
where
\beq
\Omega|_\Lambda = 0, \qquad \wt D_i \Omega|_\Lambda = 0
\eeq
and
\beq
\wt D_i \wt D_j \Omega - 2 \wt h_{ij} = 0.
\eeq
To prove this, take $\Omega = 1/r^2$ and introduce
\beq
\wt x^i = \frac{x^i}{r^2}
\eeq
as coordinates on $\wt N$. One also sees that $\wt h_{ij}$ is again the
standard flat metric in the coordinates $\wt x^i$. (This would also
follow from (3.33) and the standard formula for the behaviour of
$R_{ij}$ under conformal rescalings.) As for the potential, rewrite
(3.26) as
\beq
\left( D^2 - \frac{\R}{6} \right) \phi = 0,
\eeq
and observe that the operator in (3.35) obeys
\beq
\left( \wt D^2 - \frac{\wt \R}{6} \right) \wt \phi = 
\Omega^{-5/2} \left( D^2 - \frac{\R}{6} \right) \phi,
\eeq
when $\wt h_{ij} = \Omega^2 h_{ij}$ and $\wt \phi = \Omega^{-1/2}\phi$
for arbitrary $\Omega > 0$. Thus we again have
\beq
\left( \wt D^2 - \frac{\wt \R}{6} \right) \wt \phi = \wt D^2 \wt \phi = 0,
\eeq
at first only on $N$.

In the case of G.R. we were unable to prove convergence of the multipole
series, but only an asymptotic estimate like
\beq
\phi = \sum_{\ell=0}^{m-1} \frac{E_{i_1 \ldots i_\ell} x^{i_1} \ldots
x^{i_\ell}}{\ell! r^{2\ell+1}} + O^\infty(r^{-m+1-2\alpha}).
\eeq
But, from (3.38) for $m = 4$, it follows immediately that $\wt \phi$
extends to a $C^3$-function on $\wt N$. Thus, by continuity
\beq
\wt D^2 \wt \phi = 0 \qquad \mbox{on } \wt N.
\eeq
But it is a standard fact that solutions to the Laplace equation and, more
generally, for nonlinear elliptic systems with analytic coefficients \cite{Morrey},
are themselves analytic. Thus $\wt \phi$ has a convergent Taylor expansion
at the point $\Lambda$. But this is nothing but (3.29) in inverted
coordinates. Furthermore the multipole moments $E_{i_1 \ldots i_\ell}$
can now be viewed as the Taylor coefficients of $\wt \phi$ at $\Lambda$.
It follows from (3.39) that they have to be tracefree, and it is trivial 
that they determine $\wt \phi$ uniquely.

Suppose $\Omega$ is just required to satisfy (3.32,33). Then, given $h_{ij}$,
there is in $(\wt h_{ij},\Omega)$ the following 3-parameter gauge freedom
\beqa
\Omega' &=& \omega \Omega, \\
\wt h'_{ij} &=& \omega^2 \wt h_{ij},
\eeqa
where
\beq
\omega = (1 - b^i \wt D_i \Omega + b^i b_i \Omega)^{-1},
\eeq
with $\wt D_i b^j = 0$, which, in the compactified picture, corresponds to
the freedom of choosing an origin in the ``physical'' space ${\bf R}^3$,
w.r. to which the inversion $\wt x^i = x^i/r^2$ can be made. Therefore the
Taylor coefficients of $\wt U$ at $\Lambda$ behave under (3.40,41) in a
way which precisely corresponds to their dependence on the choice of origin.

In G.R. it is impossible to require a conformal compactification for which
(3.33) holds everywhere. We call a 3-metric $\bar h_{ij}$ on a manifold
$N \cong {\bf R}^3 \setminus B(R)$ conformally $C^k$, when there exists a
$C^k$-function $\Omega > 0$ on $N$ such that $\wt h_{ij} = \Omega^2
\bar h_{ij}$
extends to a $C^k$-metric on $\wt N = N \cup \{ \Lambda \}$ and
\beq
\Omega|_\Lambda = 0, \qquad \wt D_i \Omega|_\Lambda = 0,
\eeq
\beq
\left.(\wt D_i \wt D_j \Omega - 2 \wt h_{ij})\right|_\Lambda = 0.
\eeq
A scalar potential $\phi$ is called conformally $C^k$, when 
$\wt \phi = \Omega^{-1/2} \phi$ extends to a $C^k$-function on $\wt N$.
Given (3.43,44) there is now a much larger gauge freedom involved in
constructing the unphysical from the physical quantities, namely
\beq
\Omega' = \omega \Omega, \qquad
\wt h'_{ij} = \omega^2 \wt h_{ij}, \qquad
\wt \phi' = \omega^{-1/2} \phi
\eeq
where $\omega$ satisfies $\omega|_\Lambda = 1$. Now consider, following 
Geroch \cite{GeM}, this recursively defined set of tensor fields on $\wt N$:
\beqa
P_0 &=& \wt \phi \\
P_i &=& D_i \wt \phi \\
P_{ij} &=& TS \left[\wt D_i D_j \wt \phi - \frac{1}{2} \wt \R_{ij} 
\wt \phi \right] \\
P_{i_1 \ldots i_{m+1}} &=& TS \left[ \wt D_{i_{m+1}} P_{i_1 \ldots i_m}
- \frac{s(2s-1)}{2} \wt \R_{i_1i_2} P_{i_3 \ldots i_{m+1}}\right],
\eeqa
where $TS$ denotes the operation of taking the symmetric, trace-free part.
It turns out that the tensors
\beq
E_{i_1 \ldots i_m} = \left. P_{i_1 \ldots i_m}\right|_\Lambda
\eeq
behave under (3.45) in exactly the same way as the Newtonian moments
under the restricted gauge freedom (3.40--42) with 
$b_i = \wt D_i \omega|_\Lambda$. Thus the Ricci terms in (3.46--49)
cancel out unwanted dependencies from higher-than-first derivatives of
$\omega$ at $\Lambda$.

Now return to the expansions (3.22--25) for some fixed $m \geq 1$.
Performing, again, an inversion $\wt x^i = x^i/r^2$ and setting, in
these coordinates,
\beq
\wt \phi_M = \Omega^{-1/2} \phi_M, \qquad
\wt \phi_S = \Omega^{-1/2} \phi_S,
\eeq
\beq
\wt h_{ij} = \Omega^2 \bar h_{ij}
\eeq
with $\Omega = 1/r^2$ we find that $(\wt \phi_M,\wt \phi_S,\wt h_{ij})$
are all $C^m$. Furthermore $\Omega$ is $C^\infty$. Thus we have obtained 
a $C^m$ conformal compactification. Our proof would be complete if we could
find an elliptic system satisfied by $(\wt h_{ij}, \wt \phi_M,\wt\phi_S)$
or quantities derived from them. Doing this is not completely trivial.
We explain the essentials in the static case where $\phi_S = 0$. Thus
\beq
\bar D^2 \phi_M = 2 \bar \R \phi_M
\eeq
\beq
\bar \R_{ij} = \frac{2}{1 + 4\phi^2_M} (D_i\phi_M)(D_j\phi_M).
\eeq
Let us assume that $M \neq 0$. Define, instead of $1/r^2$ as above, a
conformal factor also called $\Omega$ by
\beq
\Omega = \frac{[(-V)^{1/2}- 1]^2}{(-V)^{1/2}}.
\eeq
It is not hard to see from (3.22--25) that this yields a 
$C^m$-compactification $(\wt \phi_M,\wt h_{ij})$ where, however, we have
for convenience replaced (3.44) by
\beq
\left.\left(\wt D_i \wt D_j \Omega - \frac{2}{M^2} \wt h_{ij}\right)
\right|_\Lambda = 0.
\eeq
It is useful to employ, as the scalar variable in the unphysical picture
neither $\wt \phi_M$ nor $\Omega$, but the quantity $\sigma$ defined by
\beq
\sigma := \left[\frac{(-V)^{1/2} - 1}{(-V)^{1/2} + 1} \right]^2.
\eeq
After some labor we find from (3.53,54) that
\beq
\wt \R = 0
\eeq
and
\beq
- \sigma(1-\sigma) \wt \R_{ij} = \wt D_i \wt D_j \sigma - \frac{1}{3}
\wt h_{ij} \wt D^2 \sigma.
\eeq
The scalar $\sigma$ satisfies
\beq
\sigma|_\Lambda = 0, \qquad \wt D_i \sigma|_\Lambda = 0, \qquad
\wt D^2 \sigma|_\Lambda = \frac{3}{2M^2} .
\eeq
Taking a ``curl'' of Equ. (3.59) we obtain
\beq
(1-\sigma)\wt D_{[i} \wt \R_{j]k} = 2(\wt D_{[i} \sigma) \wt R_{j]k}
- \wt h_{k[i} \wt R_{j]\ell} \wt D^\ell \sigma.
\eeq
If we take $\wt D^i$ of the quantity $\wt D_{[i} \wt \R_{j]k}$ and use
the Ricci and Bianchi identities we find the relation
\beq
\wt D^2 \wt \R_{jk} = \frac{1}{2} \wt D_j \wt D_k \wt \R +
2 \wt D^i \wt D_{[i} \wt \R_{j]k} 
 + 3(\wt \R_{ji} \wt \R^i{}_k - \frac{1}{2} \wt \R \wt \R_{jk})
- \frac{1}{2} \wt h_{jk} (\wt \R_{i\ell} \wt \R^{i\ell} -
\frac{1}{2} \wt \R^2).
\eeq
Using that $\wt \R$ is zero and Equ. (3.62), writing 
$\wt \R_{ij} = \tau_{ij}$, and using (3.59) to eliminate second
derivatives of $\sigma$, we obtain an equation of the form
\beq
\wt D^2 \tau_{ij} = \mbox{nonlinear terms,}
\eeq
where these nonlinear terms depend at most on $\tau_{ij}$, $\sigma$ 
and their first derivatives and on $\wt D^2 \sigma$.
We call $\wt D^2 \sigma = \rho$. From the divergence of
Equ. (3.59) we infer that
\beq
\rho \sigma = \frac{3}{2} (\wt D \sigma)^2,
\eeq
and from this after some work that
\beq
\wt D^2 \rho = 3\sigma (1-\sigma)^2 \wt \R_{ij} \wt \R^{ij} + 3 \wt \R_{ij}
(\wt D^i \sigma)(\wt D^j \sigma).
\eeq
Now Equ. (3.63) can be completed as follows:
\beqa
\wt \R_{ij} &=& \tau_{ij} \\
\wt D^2 \tau_{ij} &=& \mbox{nonlinear terms} \\
\wt D^2 \sigma &=& \rho \\
\wt D^2 \rho = 3\sigma (1-\sigma)^2 \wt \R_{ij} \wt \R^{ij} + 3 \wt \R_{ij}
(\wt D^i \sigma)(\wt D^j \sigma).
\eeqa
Going over to harmonic coordinates, the ``non-elliptic'' terms in the
expression of $\wt \R_{ij}$ in (3.66) in terms of the metric go away,
and the whole set of Equ.'s (3.66--69) becomes an elliptic system.
Note that the point of the whole man\oe uvre was that the original
Equ. (3.59), when written in terms of $\wt h_{ij}$ is singular since
$\sigma|_\Lambda = 0$. The miracle was that, in the transition from
(3.59) to (3.61) a factor $\sigma$ is obtained on both sides of
(3.61) which can be cancelled since $\sigma$ is nonzero outside
$\Lambda$ by (3.57).

Thus, taking $m$ sufficiently large and appealing to the theorem of
Morrey \cite{Morrey}, we have the

\paragraph{Theorem:} When $M \neq 0$, there is a chart in a neighbourhood
of $\Lambda$ for which $(\sigma,\wt h_{ij})$ are analytic. Consequently,
from (3.57), $\Omega$ is also analytic, and so is 
$\wt \phi_M = (1 - \sigma)^{-3/2}$.

An analogous result can be proved for a suitable set
$(\wt h_{ij},\Omega,\wt \phi_M,\wt \phi_S)$ in the stationary case \cite{BeSi2},
see also \cite{Ku}. The equations one obtains imply in particular that the
``physical'' quantities $(\bar h_{ij},\phi_M,\phi_S)$ have an analytic
chart
in a neighbourhood of each point of $N$ and thus entail the ``classic''
result of M\"uller zum Hagen on the analyticity of stationary vacuum
solutions \cite{MzHA}.

By smoothness of $(\wt h_{ij},\wt \phi_M,\wt \phi_S)$ we can define
multipole moments for each of $\wt \phi_M,\wt \phi_S$, following 
(3.46--50). One can show \cite{SiB} that they coincide with the quantities
$E \ldots$ and $F \ldots$ in the expansions (3.22--25). 
(These, in turn, coincide with the ones in Thorne \cite{Tho}, as shown 
in \cite{Gu}). One can now
prove \cite{BeSi2}, that these moments determine the stationary solution uniquely
up to isometries. We give a more careful formulation of this result only
in the static case.

\paragraph{Theorem:} Let there be two static solutions with the same
$\wt h_{ij}|_\Lambda$, the same $M \neq 0$ and the same set of
(mass-centered) multipole moments. Then the corresponding physical 
solutions $(\bar h_{ij},\phi_M)$ are isometric.

The proof is a not-too-difficult inductive argument based on (3.61,62),
(3.68,69) and (3.59,60).

There remains the question to what degree the multipole moments of stationary solutions
can be prescribed. It is fairly easy to see, e.g. from the asymptotic analysis of 
Sect.~3.1,
that the multipole moments are ``algebraically independent'', i.e. for a given finite number
 of them, there always exists a spacetime having those moments which solves the stationary
 field equations to arbitrary order in $1/r$. It is not known what conditions on moments
for high order have to be imposed in order for the multipole expansion to converge. In particular,
 convergence is not even known when only finitely many moments are non-zero.

There are of course solution-generating techniques to in principle write down the general
 stationary axisymmetric spacetime. To date the only result on existence of stationary
 asymptotically flat solutions without any further symmetry is that of Reula \cite{Re}.

We note, in passing that the above equations lend themselves to an easy proof
of a result which is often used in black-hole uniqueness
theorems (see \cite{IsrS}). Namely  an asymptotically flat, static vacuum
solution with $M \neq 0$, which is spatially conformally flat, has to be isometric
to the Schwarzschild metric near $\Lambda$. To see this, use that now the Cotton tensor of 
$\wt h_{ij}$ is zero. Thus, since $\wt \R = 0$, the left hand side of
Equ. (3.61) vanishes. Contracting the r.h. side of (3.61) with
$(\wt D^i \sigma) \wt \R^{jk}$ we find that
\beq
2(\wt \R_{ij} \wt \R^{ij})(\wt D \sigma)^2 =
(\wt \R_{ij} \wt D^j \sigma)(\wt \R^i{}_\ell \wt D^\ell \sigma).
\eeq
But, by Cauchy--Schwarz, the right-hand side of (3.70) is bounded above
by 
$$
(\wt \R_{ij} \wt \R^{ij})(\wt D \sigma)^2,
$$
which has hence to be zero.
Since $\sigma$ can not have critical points near $\Lambda$ except at
$\Lambda$ itself, it follows that
\beq
\wt \R_{ij} = 0,
\eeq
whence, from (3.59), $\wt D^2 \sigma = 3/2M^2$ and thus, in a chart
$\wt x^i$ for which $\wt h_{ij} = \delta_{ij}$ we have
$\sigma = |\wt x|^2/4M^2$, from which it easily follows that
$(\bar h_{ij},\phi_M)$ corresponds to Schwarzschild with mass $M$.

\section{Global Rotating Solutions}
\subsection{Lindblom's Theorem}
Lindblom showed in his thesis \cite{LiT}  that   stationary
asymptotically flat dissipative fluid configuration are axisymmetric. In this 
section we want  to outline and discuss this theorem.
 
There are three ingrediences of the proof:
\begin{enumerate}
\item[(i)] The local fluid field equations imply that
the fluid flow is proportional to a Killing vector $t^\mu$ provided the 
divergence of the entropy current vanishes.
  
\item[(ii)] The Killing field $t^\mu$ has an extension into the vacuum 
field of the solution.

\item[(iii)]  If the manifold of orbits of the stationary Killing vector 
$\xi^\mu$ is $R^3$ and asymptotically flat, then  $\xi^\mu$ is linearly 
independent of $t^\mu$. The two Killing fields commute and there is a 
linear combination of the two Kiliing fields which
has fixed points near which it act like  a rotation. 
\end{enumerate}

\paragraph{(i) Theorem:} Let $g_{\mu\nu}$, $T_{\mu\nu}$ be a 
stationary local solutions of the Einstein field equations for a
one--component fluid with  phenomenological heat conduction and viscosity laws 
and vanishing of the divergence of the entropy current.  Then the fluid flow is
proportional to a Killing vector.  

\paragraph{Proof:} The energy momentum tensor for a fluid with shear and bulk 
viscosity is \cite{MTW} ($\theta$ and $\sigma_{\mu\nu}$ are the expansion 
and shear of the fluid; $q^\mu$ is the heat flow \cite{EhM})

\begin{equation}
T^{\mu\nu}=\rho u^\mu u^\nu +(p-\zeta \theta) h^{\mu\nu}-
2\eta\sigma^{\mu\nu}+q^\mu u^\nu+q^\nu u^\mu
\end{equation}
with
\begin{equation}
h^{\mu\nu}=g^{\mu\nu}+u^\mu u^\nu\ , \ q_\mu u^\mu=0,\ \ \sigma_{\mu\nu}u^\mu=0 \ .
\end{equation}
This implies (a dot denotes the covariant derivative in the direction of the fluid flow $u^\mu$) 
\begin{equation}
0=-(\nabla_\mu T^{\mu\nu})u_\nu= \dot\rho+(\rho + p)\theta -\zeta\theta^2- 
2\eta\sigma^{\mu\nu}\sigma_{\mu\nu}+\nabla_\mu q^\mu +q_\mu\dot u^\mu\ . 
\end{equation}
Introducing $n$, the conserved rest--mass density, and
the specific volume $v=\frac{1}{n}$ and the specific internal energy 
$u={\rho\over n}$
we can rewrite this, using $\nabla_\mu(n u^\mu) = 0$, as 
\begin{equation}
n (\dot u+p\dot v) -\zeta\theta^2- 
2\eta\sigma^{\mu\nu}\sigma_{\mu\nu}+\nabla_\mu q^\mu +q_\mu\dot u^\mu=0\ . 
\end{equation}
For a one--component fluid we have an equation of state $u=u(p,v)$ and 
consequently there exist
scalar functions $T(p,v)$ and $s(p,v)$ with the interpretation of temperature and
specific entropy  such that
\begin{equation}
du+pdv=T ds\ .
\end{equation}
Hence $n (\dot u+p\dot v)=n T \dot s$ can be used to rewrite (4.4) as
\begin{equation}
n T \dot s -\zeta\theta^2- 
2\eta\sigma^{\mu\nu}\sigma_{\mu\nu}+\nabla_\mu q^\mu +q_\mu\dot u^\mu=0 
\end{equation}
or
\begin{equation}
n \dot s+ T^{-1}\nabla_\mu q^\mu= T^{-1}(\zeta\theta^2+ 
2\eta\sigma^{\mu\nu}\sigma_{\mu\nu} -q_\mu\dot u^\mu)=0\ . 
\end{equation}
Using again $\nabla_\mu(n u^\mu)=0$ we obtain
\begin{equation}
\nabla_\mu(n s u^\mu+T^{-1}q^\mu)= T^{-1}[\zeta\theta^2+ 
2\eta\sigma^{\mu\nu}\sigma_{\mu\nu} -q_\mu(\dot u^\mu+T^{-1}\nabla^\mu T)]=0 \ .
\end{equation}
Inserting the phenomenological law of heat conduction
\begin{equation}
q_\mu =-\kappa h^\nu{}_\mu(T_{,\nu}+T\dot u_\nu)
\end{equation}
we obtain
\begin{equation}
\nabla_\mu(n s u^\mu+T^{-1}q^\mu)= T^{-1}(\zeta\theta^2+ 
2\eta\sigma^{\mu\nu}\sigma_{\mu\nu} +\kappa T^{-1} q_\mu q^\mu)=0\ . 
\end{equation}
The left hand side of this equation is the conserved entropy current $\nabla_\mu s^\mu$ which vanishes
according to our assumptions. Hence the positivity of $ \lambda, \zeta$ and $\kappa$ implies
$\theta=\sigma^{\mu\nu}=q^\mu=0$ and
$\dot u_\mu=-T^{-1}T_{,\mu}\ .
$

Assume $T\ne 0$ and consider $\xi^\mu=T^{-1}u^\mu$, the candidate for the Killing
vector. We have
\begin{equation}
\nabla_{(\mu}\xi_{\nu)}=-T^{-2}\nabla_{(\mu} T u_{\nu)}+ T^{-1}\nabla_{(\mu}u_{\nu)}\ .
\end{equation}
The vanishing of $\theta=\sigma_{\mu\nu}=q^\mu=0$ implies
$\nabla_{(\mu}u_{\nu)}=-\dot u_{(\mu} u_{\nu)}= T^{-1}(\nabla_{(\mu}T)
u_{\nu)}$, hence $\nabla_{(\mu}\xi_{\nu)}=0$.

Now we come to the most complicated part, the extension of the Killing
vector proportional to the fluid flow from the fluid into the surrounding vacuum region. 

\paragraph{(ii) Conjecture: }Let $g_{\mu\nu}, T_{\mu\nu}$ be a strictly stationary, asymptotically flat
 perfect fluid solution where the matter is a ball of finite extent and 
the fluid flow is proportional to a Killing vector $t^\mu$.
 Then  $t^\mu$ has  a unique extension into the vacuum region, provided certain
differentiability conditions are satisfied at the boundary. 

In Lindblom's original
treatment this conjecture was shown to be true under the assuption that the outside metric
is analytic up to and including the boundary $\Sigma$ of the fluid. Then one
can propagate the Killing vector into a neighbourhood of the boundary using the Cauchy
Kowalevskaja theorem because a Killing vector satisfies a wave type
equation. Finally a
theorem by Nomizu \cite{No} can be used to obtain a global Killing 
vector field. 

One might, however argue, that analyticity up to and including $\Sigma$ is too strong
an assumption. On physical grounds one would like to treat also non-- analytic equations
of state. In this case it is unlikely that the metric is analytic in the boundary. 

Finally we  show that the new Killing vector $t^\mu$ is actually different from the 
stationary Killing vector $\xi^\mu$. 

\paragraph{(iii) Theorem:} Under the assumption of the above conjecture 
we have:

(1) The Killing vectors $\xi^\mu$ and $t^\mu$ are linearly independent.

(2) Both Killing vectors commute.

(3) There exists a linear combination $\eta^\mu =t^\mu + a\xi^\mu$ which has fixed points and acts like a 
rotation with closed orbits.

\paragraph{Proof:}(1) Suppose $t^\mu$ would be linearly dependent of
$\xi^\mu$. Then there would be a timelike Killing vector, namely $t^\mu$,
which is asymptotically a translation and relative to which the matter
does not rotate. Hence, by the Licherowicz staticity theorem, spacetime
would have to be static.

(2) As $T$ and $u^\mu$ are invariant objects we have ${\cL}_{\xi} T=0 $ 
and ${\cL}_{\xi}u^\nu=0 $ which  imply immediately $[\xi,t]=0$ 
on the support of the matter. 
To show that this is also true outside the matter one can use the analyticity 
of the outer metric up to and 
including the boundary or one can use a theorem by Beig and 
Chrusciel \cite{BeChr2} classifying all possible group action
on asymptotically flat spacetimes.  

(3) As $\xi^\mu$ commutes with $t^\mu$ there is a Killing vector $\hat t^i$ on the
manifold of orbits of $\xi^\mu$. The corresponding  group acts in the 2--surface
of constant pressure, in particular in the boundary, $p=0$. As this is topologically
$S^2$, there must be a point where $\hat t^i$ vanishes. A Killing vector on a
Riemannian space with a fixed point acts always as a rotation with closed circular
orbits. At a point $q$ in spacetime projecting on the fixed point of $\hat t^i$, 
$t^\mu$ must be proportional
 to $\xi^\mu$ and therefore a linear combination
$\eta^\mu=t^\mu+a \xi^\mu$
 with constant coefficients vanishing at $q$ exist such that $\eta^\mu(q)=0$. We have a fixed point
 and because
 also the timelike direction of $\xi^\mu$ is fixed,
  $\eta^\mu$ acts like a rotation  and  has therefore closed orbits. 

We see that we can obtain the existence of the axis working only on the body,
provided we know that both Killing vectors are independent. Lindblom 
\cite{LiA} obtains the
axis and commutativity of the Killing vectors from the asymptotic symmetry group. The key
property that the two Killing vectors are linearly independent is only implied by a global
argument and uses asymptotic flatness.

\subsection{Existence of Stationary Rotating Axi-Symmetric Fluid Bodies}

Following work by Liapunoff and Poincar\'e, Lichtenstein \cite{Licht} demonstrated at the beginning
of this century the existence of  rotating fluid bodies in Newtonian theory.  An account of
this almost forgotten work can be found in \cite{SchmL}. Using implicit function theorem techniques
--- as we would say today --- he shows the existence of solutions near known solutions or
approximately known solutions: starting with a static fluid ball, he obtains a slowly
rotating fluid ball; starting from a self gravitating 2--body point particle solution, he
obtains a solution for two small fluid bodies orbiting their center of mass on a circle.
Furtheremore, there is a number of exact solutions in Newtonian theory: the Maclaurin
spheroids, the Jacobi and the Dedekind ellipsoids and the Riemann ellipsoids
\cite{Chandra}.

In Einstein's theory we do not know any  stationary exact solution describing an extended
rotating body. Spacetimes describing  such solutions can be characterized as follows:
Besides a timelike Killing vector $\xi^\mu$ there is a further symmetry, the axial symmetry generated by $\eta^\mu$, whose
orbits are circles (Remember that we showed  in Sect.~4.1 that such an extra symmetry
exists on physical grounds) The body is spatially compact and the spacetime with topology
$R^4$ is assumed to be asymptotically flat. We assume that there is an axis where $\eta^\mu$ vanishes.
Then we can use a result of Carter \cite{CaCom}  which states that under these
circumstances the two Killing vectors commute. Such spacetimes are called 
"stationary axisymmetric", the orbits of the axial Killing vector are circles.

We showed in Sect~2.2 that for stationary axisymmetric  perfect fluids with
an axis and a fluid flow vector contained in the two-surface spanned by the two Killing vectors, the 
two-surface
 elements orthogonal to the two-dimensional group orbit are
surface forming ( the group action is orthogonally transitive). 
The same holds in the vacuum region.
The property of orthogonal transitivity is equivalent to the existence of 
a discrete isotropy group \cite{SchSph}.   
  
To introduce a global coordinate system let us assume  that outside the 
2-dimensional axis
the spacetime is the product of the orbits of the isometry group and the orthogonal 2 --
surface which we assume to have topology $R^2$. 

Using coordinate adapted to the Killing vectors the metric can be written as 
\beq
ds^2=g_{AB}(x^c)dx^A dx^B +
g_{00}(x^C)dt^2 +2 g_{0\phi}(x^C) dt d\phi + g_{\phi\phi}(x^C)d\phi^2 .
\eeq

Locally we can always introduce coordinates $(x^A)=(r,z)$ such that  $g_{AB}$ is conformal
to the flat metric in standard coordinates and can therefore write the metric as
\beq
ds^2=e^{2k-2U}(dr^2+ dz^2)+
e^{-2U}W^2 d\phi^2 - e^{2U}(dt+Ad\phi)^2 . 
\eeq

There is the freedom in $(r,z)$ of an arbitrary conformal transformation which is given by
the real part of analytic function.

The function $W^2$ is the volume element of the group orbits. As a consequence of the field
equations in vacuum one can locally achieve $W=r'$ such that 
\beq
ds^2=e^{2k'-2U}(dr'^2+ dz'^2)+
e^{-2U}r'^2 d\phi^2 - e^{2U}(dt+Ad\phi)^2 .  
\eeq
These coordinates are called Weyl's canonical coordinates. Matters can be arranged 
so that
$r'=0$ is the axis. Then the coordinates are fixed up to a translation in 
$z'$.

It is tempting to try to extend the Weyl coordinates from the outside of the body to the
interior such that the two-surface orthogonal to the group orbit is covered by one
$(r',z')$ system with $r'=0$ describing the axis and $W\ne r'$ in the interior.
However, M\"uller zum Hagen has demonstrated \cite{MzHaSph} that this is 
impossible in the case of
static spherically symmetric solutions. ($r'$ becomes negative inside the body and the
axis is reached for $\rho'\to\infty$.) There is no reason to assume that 
this would be different in the stationary case. 

Numerical codes work successfully with a global $(r,z)$ coordinate system such that $r=0$
is the axis but it is not assumed that one has Weyl's canonical coordinate in  
vacuum.  

For perfect fluids whose velocity is proportional to a constant linear combination of the
two Killing vectors, the case of rigid rotation, 
$\nabla_\nu T^{\mu\nu}{}=0$ becomes
particularly simple. (See equation (2.69).)
\beq
0= \nabla_\nu T_\mu{}^\nu =(\rho +p){1\over 2}(\ln f^{-2})_{,\mu} + 
p_{,\mu} .
\eeq
where $f^\mu=f^2(\xi^\mu + \Omega\eta^\mu) $ is the four velocity of the fluid.
This shows that, provided an equation of state $\rho(p)$ is given, 
the matter variables $p$
and $\rho$  can be expressed as functions of the quantity $f$ which is determined by the
geometry. This property of rigidly rotating fluids is essential for all the numerical 
schemes as well as for all the attempts to prove existence. 

Various authors have developed codes to calculate numerically stationary, axisymmetric
rigidly rotating fluid solutions \cite{Num}. Today this can be done with very high 
presicion by
different numerical techniques. These numerical solutions are also the basis for
investigations  of oscillations of rotating stars.

Schaudt and Pfister \cite{SchPf} try to obtain an existence theorem working in the above
coordinates adapted to the symmetry. This approach is attractive because the field
equations become semilinear elliptic. One has, however, to control the 
singularities in
the equations on the axis. This is possible and two Dirichlet problems have been solved,
which give existence of outside, asymptotically flat solutions and existence of inner parts
of bodies, provided appropriate boundary values are given \cite{Scha}. 
Up to now this was only possible
for the  "outer" and the "inner" problem separately and work is in progress which tries to
combine the inner and outer solution.

Let us now turn to the discussion of the only existence theorem for rotating fluids in
Einstein's theory. It is remarkable that the first existence theorem for rotating fluids,
proved by Heilig in 1995 \cite{HeGr}, uses Lichtenstein's  technique and does not adapt the
coordinate to the axial Killing vector to avoid difficulties at the axis. 

Let us formulate one particular case of the theorem proved by Heilig 
\cite{HeGr}.

\paragraph{Theorem:} Let $\rho(p)=Cp^\gamma$ be a polytropic equation of state with
$1<\gamma<6/5$.
 The central density $\rho_0$ determines a unique Newtonian static fluid 
ball solution  of finite extent.
Then there exist a  positive constant  $\Omega_0$ such that for all  $\Omega$ with 
$0<\Omega<\Omega_0$
 a stationary
axisymmetric rigidly rotating  solution with angular velocity $\Omega$  of the Einstein
 field equations for a perfect fluid
exists. The solution is geodesically complete, asymptotically flat with  finite mass
 and
angular momentum. The matter is of finite extent and has the same equation of state 
and central density as the Newtonian solution.

The theorem  holds also for more general equations of state. It is not clear whether the
case of positive boundary density may be treated by this method.

 Heilig  uses the observation of J\"urgen Ehlers \cite{EhN} that
it is possible to write the field equations as an elliptic system with a parameter
$\lambda=c^{-2}$ --- interpreted as the velocity of light --- such that the equations  for
$,\lambda\to 0$ give the Newtonian equations and the limit is regular. This can be achieved
by a particular choice of unknowns for which the field equations are formulated. 

We will describe the structure of Heiligs proof using the equations formulated 
in Sect.~2.4 because these are much simpler. We want, however, to stress 
that we expect that Heilig's result could be proved more easily using 
these equations, but this is not certain 
before  all the functional analysis has been done properly.

Let us first adapt the field equations to a rigidly rotating fluid. We write 
the axial Killing vector as in (2.54) 
\begin{equation}
\eta_\mu dx^\mu=\eta(cdt +\phi_idx^i) + \eta_idx^i\ . 
\end{equation}
The Killing equation in spacetime is equivalent to the equations (2.57--59) 
on the quotient $N$. 
 For a rigidly rotating perfect fluid with  fluid flow vector $u^\mu$ 
we have
\begin{equation}
u_\mu=f(\xi_\mu+\Omega\eta_\mu), \ \ \  \Omega={\rm const}\ , \ \ \  
u_\mu u^\mu =-c^2,
\end{equation}
where
\begin{equation}
f^{-2}=e^{-\frac{2U}{c^2}}(-e^{\frac{2U}{c^2}}+c^{-1}\Omega\eta)^2
-c^{-2}\Omega^2\eta_l\eta^l\ .
\end{equation}
To obtain the field equation we replace in (2.64--66) $\eta$ by $c^{-1}\eta$ and $p$ by
$c^{-2}p$ to obtain from (2.46--48)  using $U\rightarrow c^{-2}U$   
\beqa
\bar D^2U &=& 4\pi
G[f^2(- e^{\frac{2U}{c^2}}+ c^{-1}\Omega\eta)^2(\rho+c^{-2}p)
+2c^{-2}p e^{-\frac{2U}{c^2}}] \no \\
&& \mbox{} +c^{-2} e^{\frac{4U}{c^2}} f^2\Omega^2(\rho+c^{-2}p)
\eta_i\eta_j\bar h^{ij}] e^{\frac{4U}{c^2}}
-e^{\frac{4U}{c^2}}\bar\omega_{ij}\bar\omega^{ij} \\
\bar D^i\bar\omega_{ij} &=& 8\pi G
c^{-3}e^{-\frac{4U}{c^2}}
f^2(-e^\frac{2U}{c^2}+c^{-1}\Omega\eta)(\rho+c^{-2}p)\Omega\eta_j \\
\bar R_{ij} &=& 2c^{-4}D_iUD_jU 
-2e^{\frac{4U}{c^2}}\bar\omega_{ik}\bar\omega_j{}^k+
\bar h_{ij}e^{\frac{4U}{c^2}}\bar\omega_{kl}\bar\omega^{kl} \no \\
&& \mbox{} +
8\pi G c^{-4}[
-2pe^{-\frac{2U}{c^2}}\bar h_{ij}+f^2\Omega^2(\rho+c^{-2}p)\eta_i\eta_j
\no \\
&& \mbox{}-\bar h_{ij}f^2\Omega^2(\rho+c^{-2}p)\eta_l\eta_m\bar h^{lm}]
\eeqa
The above field equations have to be supplemented by the Killing equations (2.57--2.59).
The equations of motion are
\begin{equation}
\nabla_\mu T^{\mu\nu}=0\Longleftrightarrow (c^2\rho +p)f^{-1}D_if = D_ip 
\end{equation}

For $c\to\infty$ we have from (4.18) that $f^2=1$ which implies by (4.20)  
that $\bar D^i\omega_{ij}=0$. The staticity theorem now gives that 
$\bar\omega_{ij}=0$. Using (2.58) this implies
$D_i(e^{-\frac{2U}{c^2}}\eta)=0$. The vanishing of $\eta$ on the axis implies
$\eta=0$. Using all this the field equations reduce to 
\begin{equation}
\bar R_{ij}=0\ ,\ \ \ \bar D^2 U=4\pi G \rho
\end{equation}
Therefore the metric on $N$ is flat. Using
$lim_{c\to\infty}(c^2D_if)=-D_i(U-\frac{1}{2}\Omega^2\eta_l\eta^l)$
the equation of motion become the Newtonian equation
\begin{equation}
-\rho D_i(U-\frac{1}{2}\Omega^2\eta_l\eta^l)=D_ip
\end{equation}
equation ($\eta_l\eta^l=x^2+y^2$ in Cartesian coordinates).

As discussed in Sect. 2.5 for the static case, the field equations become again a quasilinear
elliptic system for $U, Z^{ij}, \vp_i$ in harmonic coordinates ($\nabla_\mu\nabla^\mu
t=0\Longleftrightarrow \bar D^i\vp_i=0$,  $\nabla_\mu\nabla^\mu
x^i=0\Longleftrightarrow \bar D^2 x^i=0$).
Namely, the condition that the time function is harmonic turns the left-hand side of Equ.(4.20)
into an elliptic operator acting on $\vp_i$. Harmonicity of $x^i$ has the same effect on the
 left hand side of Equ.(4.21).
 Theorem~4.1 of Heilig 
\cite{HeGr} can be
adapted to show that for small 
$\lambda,\Omega$
a solution of the reduced field equation exists near the Newtonian solution. Such a solution
satisfies only the harmonicity condition if the equation of motion holds. So, this has to be
solved simultaneously. This is possible because given a equation of state 
(4.22) can be
integrated such that the matter quantities can be expressed in terms of 
the geometrical quantity $f$. Therefore the following iteration procedure 
is well defined: begin with a Newtonian solution
$U^0,p^0$; choose some $\lambda, \Omega$ and  use $\rho^0,p^0,U^0$ as a source in the field
equations in harmonic coordinates to obtain $U^1,Z^1{}^{ij},\vp^1{}_{i}$. Calculate
$f$ from $U^1,Z^1{}^{ij},\vp^1{}_{i}, \lambda,\Omega$ and determine $p^1$ from the
equation of motion. Then one solves again the field equation with the new source and so
on. Heilig has shown that for sufficiently small $\lambda$ and $\Omega$ such an
iteration converges in his variables. It should also converge in the variables used
here.

It is remarkable that we have used $\eta_idx^i=xdy-ydx$ as a given field. At the end one has
 to check that
the solution is axisymmetric and satisfies the harmonicity condition. 

Note that only for $\lambda=c^{-2}$ with some fixed value of the velocity of light in some units the
above field equations are Einstein's equations. It is however possible to reinterpret 
 solutions
with any $\lambda$ as solutions of Einstein's equation expressed in different units
 \cite{HeGr}.
With this interpretation the theorem above demonstates the existence of slowly --- the
theorem does not control the range of $\omega$ --- rotating fluid configurations. 

\subsection{The Neugebauer--Meinel Disk}
The only known global solution describing a rotating object in Einstein's
theory, is the relativistic analogue of the rigidly rotating Maclaurin
disk in Newton's theory \cite{BiTr}. 

An axisymmetric surface density distribution (in cylindrical coordinates
$(r,\phi,z )$)
\beq
\sigma(r)=\sigma_0\sqrt{1-{r^2\over r^2_0}}\ ,\ \ 0<r<a\ ,
\eeq
generates a gravitational potential $\Phi(r,z )$, which is determined
by the Poisson integral from $\sigma$. At the disk the potential is
\beq
\Phi(r,0)={1\over 2}\Omega^2r^2 +\rm{const}\ , \ \ 0<r<a\ ,\ \ 
\Omega^2={\pi^2G\sigma_0^2\over r_0^2}\ .
\eeq
Outside the disk the potential can be expressed, for example, in terms of integrals
over Bessel functions.

The centifugal force acting on rigidly rotating  particles balances the
gravitational force in the disk. Therefore, we can interpret the density
distribution as formed by  self gravitating, rigidly rotating dust. The
two parameters $\sigma_0$ and $r_0$ determine such disks uniquely. The total
mass of the disk is $M={2\over 3}\pi \sigma_0 r_0^2$.

Neugebauer and Meinel found the relativistic analog of these disks 
\cite{NeuMei}.

There is a well known formalism available in General Relativity to describe matter
surface distributions \cite{IsrS}. In the particular case of a
reflection symmetric disk, we have to find solutions of the stationary
vacuum field equations, defined outside the disk such that the difference
of the normal derivatives of the metric at of the disk have a certain
algebraic structure \cite{IsrS}.

The general stationary axisymmetric metric can be parametrized as 
\beq
ds^2= e^{-2U}\left[e^{2k}(dr^2+dz^2)+r^2 d\phi^2\right] - 
e^{2U}(dt+ad\phi)^2 .
\eeq
The metric coefficients $U,k$ and $a$ depend only on $r,z$; the vector
fields
$\xi^\mu{\partial/\partial x^\mu}=\partial/\partial t$ and 
$\eta^\mu{\partial/\partial x^\mu}={\partial/\partial\phi}$ are
Killing vector fields. We assume that the orbits of the axial Killing
vector are circles; $r=0$ is the  axis. 

Let us assume that the disk is located at
$z=0,\ 0\le r<r_0$. A rigidly rotating flow forming the disk is described by a
vector field (which is defined at the disk)
\beq
u^\mu=e^{-V}(\xi^\mu + \Omega \eta^\mu)\ , \ \ u^\mu
u_\mu=-1\ ,
\eeq
where $\Omega$ is constant. The definition of a dust disk implies that the 
metric is continuous across the disk and that $\tau^{\mu\nu}=\sigma
u^\mu u^\nu$, where $\sigma$ is the surface density, satisfies
$\tau^{\mu\nu}{}_{;\nu}=0$ with respect to the Levi Civita connection
of the metric induced on the disk. As $v^\mu=\xi^\mu + \Omega \eta^\mu$ is
a Killing vector, it holds $v^\mu{}_{;\mu}=0, \ \sigma_{,\mu} v^\mu=0$,
$V_{,\mu}v^\mu=0$ and we obtain
\beq
\tau^{\mu\nu}{}_{;\nu}=\left(\sigma e^{-2V}v^\mu v^\nu\right)_{;\nu}
=\sigma e^{-2V}v^\mu{}_{;\nu} v^\nu .
\eeq
Finally $e^{2V}=g_{\mu\nu}v^\mu v^\nu$ implies
$e^{2V}2V_{,\gamma}=2g_{\mu\nu}v^\mu v^\nu{}_{;\gamma}=
2v^\nu v_\gamma{}_{;\nu}$ and we see that $V$ must be constant on a disk 
formed of rigidly rotating dust, $V=V_0$.
  
It is natural to introduce comoving coordinates
\beq
t'=t;\ \ \phi'=\phi-\Omega t \ ,\ \  \xi^{\mu'}=\xi^\mu+
\Omega\eta^\mu\ ,\ \ \eta^{\mu'}=\eta^\mu\ ,\ 
u^{\mu'}=e^{-V}\delta^{\mu'}_{t'} .
\eeq
The vacuum field equations can be expressed in terms of the following
quantities:
\beq
e^{2U'}=-\xi_{\mu'}\xi^{\mu'}=e^{2V}\ ,\ \
a'=-e^{-2U'}\eta_{\mu'}\xi^{\mu'}\ ,\ 
U'(r,\phi,z=0)=V_0=const
\eeq
and $b'(r,z)$ determined by
\beq
a'_{,r}=r e^{-4U'}b'_{,z}\ , \ \ a'_{,z}=-r e^{-4U'}b'_{,r}\ .
\eeq
Using the Ernst potential $f'=e^{2U'}+ib'$  the key field equation is the
semilinear elliptic Ernst equation  \cite{ExSol}
\beq
Re(f')(f'_{,rr}+f'_{,zz}+{1\over r}f'_{,r})
=f'^2_{,r}+f'^2_{,z}\ .
\eeq
For a solution of the Ernst equation the integrability condition of (4.32) is
satisfied and one can solve for $a'$. The remaining metric coefficient $k'$
follows from the equations 
\beq
k'_{,r}=r\left[U^{'2}_{,r}-U^{'2}_{,z}+{1\over
4}e^{-4U'}(b^{'2}_{,r}-b^{'2}_{,z})\right]\ ,\ \ 
k'_{,z}=2r\left[U'_{,r}U'_{,z}+{1\over
4}e^{-4U'}(b'_{,r}b'_{,z})\right]\ ,
\eeq
whose integrability condition is again satisfied for solutions of the
Ernst equations.

We can perform an integral in the $(r-z)$ --plane around the disk of the
integrability condition of (4.32), namely 
\beq
(r^{-1}e^{4U'}a'_{,r})_{,r}+(r^{-1}e^{4U'}a'_{,z})_{,z}=0\ ,
\eeq
which can be replaced by a surface integral.
As we assume that the tangential derivatives of the metric are continuous at the disk, we
obtain at the disk
\beq
a'_{,z}|_{z=0^+}=a'_{,z}|_{z=0^-}\ .
\eeq
On the other hand reflection symmetry at the disk implies 
\beq
a'_{,z}|_{z=0^+}=-a'_{,z}|_{z=0^-}
\eeq
on the disk, hence, 
\beq
a'_{,z}|_{z=0^+}=a'_{,z}|_{z=0^-}=0\ ,
\eeq
which by (4.32) implies $b'=const$ on the disk.  

Now it is easy to calculate the second fundamental form
$k_{cd}={1\over 2}e^U g_{cd,z}$ of the disk $z=0$ ( $c,d,\dots
=(t,r,\phi)$),
because we have at the disk that $a'_{,z}=k'_{,z}=0$, as a consequence of
(4.38),(4.35) and (4.34). We find
\beq
k_{rr}=-2U'_{,z}g_{rr}
\eeq
\beq
k_{\phi'\phi'}=-2U'_{,z}(a'^2e^{2U'}+e^{-2U'}r^2)
\eeq
\beq
k_{t't'}=2U'_{,z}g_{t't'}
\eeq
\beq
k_{t'\phi'}=2U'_{,z}g_{t'\phi'}.
\eeq

Now we can check the condition for a disk of dust \cite{IsrS}, namely
\beq
^+k_{cd}-{}^-k_{cd}=2^+k_{cd}=-8\pi(\tau_{cd}-{1\over2}g_{cd}\tau^e_e)=
-8\pi(\sigma u_cu_d+{1\over 2}\sigma g_{cd})
\ ,
\eeq
which, in the primed coordinates (because of $u^{\mu'}=\delta^{\mu'}_{t'}$),
reads
\beq
k_{c'd'}=-8\pi\sigma(g_{c't'}g_{d't'}+g_{c'd'})\ .
\eeq
Because of the form of the metric (4.27) in primed coordinates and by
(4.39--42), this is satisfied if we define the surface density by
\beq
\sigma={1\over 2\pi}U'_{,z}\ .
\eeq
Thus we have shown that  a rigidly rotating disk of dust is determined by a solution of the Ernst
equation which satisfied at the disk $U'=const$ and $b'=const$. Outside the disk the
solutions of the Ernst equation must be regular. For a well-posed elliptic problem
we need furthermore asymptotic flatness at infinity and regularity conditions at
the axis.

In Newton's theory  there is  a 2--parameter family of disks (4.25), (4.26). If we use
the 2--parameter group of similarity transformations --- or dimensionless quantities
--- we can assume $r_0=1$ and $\sigma_0=1$ and we have just one disk.

Because of the appearence of the velocity of light there is only a 1--parameter
group of similarity transformations in Einstein's theory. Hence, after we put $r_0=1$, we expect a
1--parameter family of disk solutions. 

The investigations of Neugebauer and Meinel suggest that
\beq
\mu=2\omega^2r_0^2e^{-2V_0}
\eeq
is an appropriate parameter.

Neugebauer and Meinel prove by the so called inverse scattering method of soliton
physics (compare the
contribution of Maison in this volume) that the boundary value problem for $f'$ has a unique global
solution provided $V_0$ and $\omega$ are such that $\mu<\mu_{crit}=4.629\dots$.  

The solution $f'$ can be expressed in terms of hyperelliptic theta functions \cite{NeuMei1}.
The remaining metric coefficients $a'$ and $k'$ are determined by integration from
the equations (4.32) and (4.34).

If we put  the velocity of light, $c$,  in the appropriate places we obtain the
MacLaurin disk as a Newtonian limit.

Further properties of these disks are discussed in \cite{NeuMei1}.

Many global stationary solutions with disk sources may be constructed from known
stationary vacuum solutions by ``cutting out'' a region containing singularities an
making appropriate identifications. This method was first used by Bicak and
Ledvinka \cite{BiL}
to produce physically plausible sources for the Kerr metric with arbitrary values of the 
parameters $a,M$. These disks are made of two streams of particles circulating in opposite
 directions with differential velocities. They are extending to infinity but have finite mass.
 See Sect.~6 of the article by Bicak in this volume, where this procedure 
is related back to the 
``method of images'' in Newtonian galactic dynamics. In the static case
these methods yield an infinite number of such static disk  solutions.
Solutions corresponding to stationary counterrotating dust disks of finite
extent have been constructed by Klein and Richter \cite{KlR}.

\section{Global Nonrotating Solutions}

\subsection{Elastic Static Bodies}

No doubt, Einstein's theory should allow for the description of static, solid
bodies. It is useful to make the following distinction: 

(i) small bodies, whose
shape is not dominated by gravitational forces, like a piece of sugar 
or an iron ball. If we ignore gravity, the structure of  the body is 
determined by the
laws of quantum mechanics. This is true in a Galilei invariant formulation as
well as in a special relativistic one. Linear and nonlinear elasticity theory
describes the deformation of such a configuration under external forces.

Suppose we now want to add the gravitational field. This is straightforward for linear
elasticity in Newtonian theory; we just have to insert the gravitational field calculated from
the Poisson integral as an external force  into the equations of elasticity.

To pass from special relativity to Einstein's theory  is more complicated. Now
the deformed configuration has to satisfy Einstein's field equations, and the 
elasticity equations are a consequence of the latter!

(ii) bodies like stars whose shape is dominated by gravitational forces. 
There a
relaxed state does not really exist and one has to modify the desciption of
elasticity. This holds in Newtonian theory as well as in Einstein' theory. 

Elastostatics can be described in Einstein's theory as follows 
\cite{CaQ}. 
The collection of particles which form the body is described by the 
three-dimensional ``body manifold'' $B$. The
essential dynamical variable is a map $ \Phi:M^4\to B$. such that
$\Phi^{-1}(y^i)$ is the world line of the particle  in spacetime labeled by $y^i$ in
 $B$.
In the static case the world lines of the particles are the integral curves of
the Killing vector, and we can consider $\Phi$ as a 1--1  map  $N \to B$. We assume
that we have given on $B$ a Riemannian metric $\bar\kappa_{ij}$. Its physical
interpretation may be different: for small bodies it describes a relaxed
state; for big bodies which go never into a relaxed state, it could be an 
``isotropic state of minimal energy''.

We need now information about the energy momentum tensor of the material. Let
\begin{equation}
T^{\mu\nu}=\rho u^\mu u^\nu + p^{\mu\nu}\ , \ \ p^{\mu\nu}u_\nu=0
\end{equation} 
be such that the stress tensor $p^{\mu\nu}$ has only spatial components  and can
be considered as a tensor on $N$. We can now define 
\begin{equation}
e_{ij}:= {1\over 2}(h_{ij} - \Phi_*\bar\kappa_{ij})
\end{equation}
the ``Lagrangian strain tensor''. In the Hookian approximation of elastcity one
assumes that one has given  on the body  $B$ a tensor field $\bar K^{ijk\ell}$  such that after
 moving this object with $\Phi$ into the space $N'$ one can
define
\begin{equation}
\rho=\rho_0 + {1\over 2}K^{ijk\ell}e_{ij}e_{k\ell} 
\end{equation}
\begin{equation}
p^{ij}=-K^{ijk\ell}e_{k\ell}
\end{equation}
as the energy and stresses of the body $B$ in  3--space with the metric $h_{ij}$.

With this energy momentum tensor we consider Einstein's field equations
as differential equations for the spacetime  metric and the map $\Phi$. No general existence
theorem is available for this problem. The only case treated so far is the spherically 
symmetic one \cite{Park}.

To get some feeling for these equation let us consider some further idealisation. For small
deformations we can linearize   
$e_{ij}:= {1\over 2}(h_{ij} - \Phi_*\bar\kappa_{ij})$ as follows: Suppose
that
$\Psi_\epsilon$ is a 1--parameter family  of diffeomorphism $N'\to N'$
 such that  $\epsilon =0$ is the identity and $\Phi_0$ is some diffeomorphism $N'\to B$. Now we assume 
that $\Psi_\epsilon \Phi^{-1}{_0}$ defines our deformed body and calculate the stress tensor $e_{ij}$ to
 first order in $\epsilon$  
If we define $\kappa_{ij}^0=\Phi_{0*}\bar\kappa_{ij}$ we obtain
\begin{equation}
e_{ij}={1\over 2}(h_{ij}-\kappa^0_{ij}+ \cL_\chi\kappa^0_{ij})={1\over
2}(h_{ij}-\kappa^0_{ij}+ D^0_{(i}\chi^k\kappa^0_{j)k})
\end{equation}
Here the vector field $\chi^i$ is defined by the linearization of $\Psi_\epsilon$ on $N'$ . We
see that $p^{ij}{}_{;j}=0$ leads to second order differential equations for $\chi^a$.

Consider first  the case of special relativity which coincides with Galilei invariant
 classical mechanics in the static situation. Then we have 
$h_{\alpha\beta}=\eta_{\alpha\beta}$ and $h_{ij}=\delta_{ij}$. We choose $\Phi_0$ to be the
identity may which discribes the relaxed body in spacetime. With
$\bar\kappa_{ij} =\delta_{ij}$ we obtain
\begin{equation}
e_{ij} ={1\over 2}\chi_{(i,j)}
\end{equation}
This gives implies the equations of classical, linearized elastostatics
 \cite{MaHu}. 
\begin{equation}
0=p^{ij}{}_{,j}=-K^{ijk\ell}\chi_{k,\ell j}
\end{equation}
With the
 appropriate symmetry and positivity conditions on $K^{ijk\ell}$ the 
equations are elliptic  and solutions  exist for various boundary conditions.

Next we want to calculate the deformation of a small elastic body by its own gravitational 
field. The relaxed state is determined by solid state physics as above. To switch on gravity
we assume that we have families $g_{\mu\nu}=\eta_{\mu\nu}+G g^1_{\mu\nu}+G^2 g^2_{\mu\nu}\dots $
and $T_{\mu\nu}=T^0_{\mu\nu}+G T^1_{\mu\nu}+G^2 T^2_{\mu\nu}\dots $ satisfying the field
equations. 

At order $G^0$ we obtain the trivial solution if there are no forces at the body ,
the density $\rho^0_0$ is constant and the stresses vanish,i.e. $\chi^0_a=0$. The field equation
in order
$G$ are obtained from the equations in section 2.4 with an energy momentum tensor
$T^0_{\mu\nu}$ which has only a term $\rho^0_0$ because the stresses vanish. We obtain $U^1$ as a
solution of the Poisson equation with the source  source $\rho^0_0$. The metric $\bar h^1_{ij}$
remains flat in this order. The expansion of the equation of motion in $G$ gives to first
order 
\begin{equation}
p^1{}^{ij}{}_{,j}=-K^{ijk\ell}\chi^1_{k,\ell j}=-\rho^0_0 U^1{,_a}\ , 
\ \ \ \  \Delta U^1=4\pi G\rho^0_0 
\end{equation}
Hence we obtain classical elastostatics with the force deforming the body
being the gravitational force.

One might try to obtain an existence theorem for small self gravitating elastic bodies in
Einstein's theory by an implicit function theorem argument similarly as in the case of a
rigidly rotating body (section 4.2). 

\subsection{Are Perfect Fluids $O(3)$-Symmetric?}
It is intuitively ``obvious'' that a static, in particular non-rotating,
ball of perfect fluid, due to the absence of shear stresses should have
spherical symmetry, and in particular the gravitational field in its
exterior should be the one described by the Schwarzschild spacetime.
This result, in its most general form, is still open in G.R. (The 
Newtonian case was settled in Lichtenstein \cite{Licht} and Carleman
\cite{Carl},
see
also Lindblom \cite{LiN}.) Rather, one has today a theorem which is
essentially
a uniqueness result in the spirit of black hole uniqueness theorems. 
An earlier result due to K\"unzle and Savage \cite{KuSaU} states that,
near a
spherical solution, there is no aspherical one with the same equation
of state and the same mass.

The following result, due to Beig and Simon \cite{BeSiF2}, is a refinement
of previous work by Masood-ul-Alam \cite{MuA}, see also the review of
Lindblom \cite{LiSy}.
 
\paragraph{Theorem :} Let us have a static, asymptotically flat,
spherically symmetric solution to the Einstein equations with a perfect
fluid and barotropic equation of state $\rho = \rho(p)$. (This solution
is called reference spherical solution.) Let there be given another static,
asymptotically flat solution with the same equation of state and the same
value $V|_{\partial S}$ of the Killing vector norm on the surface
$\partial S$ of the star. Let further $\rho(p)$ satisfy the differential
inequality $I \leq 0$, specified later. Then these two spacetimes are
isometric, in particular the second one is also $O(3)$-symmetric.

The condition stipulating the existence of a spherical reference solution
was disposed of by Lindblom and Masood-ul-Alam \cite{LiMuA}. The condition
on 
the matter, besides the one stating that $\rho \geq 0$, $p \geq 0$ and
$d\rho/dp \geq 0$, is that
\beq
I := \frac{1}{5} \kappa^2 + 2 \kappa + (\rho + p) \frac{d\kappa}{dp}
\leq 0,
\eeq
where $\kappa := \frac{\rho + p}{\rho + 3p}\; \frac{d\rho}{dp}$. One can
check that it is for example satisfied for the equation of state of a
relativistic ideal Fermi gas at zero temperature, but only up to
densities of order $10^{15}$gcm$^{-3}$, which is roughly the critical
density where gravitational instability sets in. It is known from numerical
results \cite{WSchm} that beyond that limit the uniqueness statement of
the above
theorem will fail. One believes however, that sphericity will still hold.

We will here confine ourselves to an outline of the proof to the case of
the special equation of state given by \cite{Bu1}
\beq
\rho(p) = \frac{1}{6} \; \rho^{6/5}(\rho_0^{1/5} - \rho^{1/5})^{-1}
\qquad (\rho_0 = \mbox{const} > 0)
\eeq
which is a relativistic generalization of the equation for a polytrope 
of index 5 in the Newtonian theory . The expression in (5.10) satisfies
$I \equiv 0$. The reference spherical solution in this case is known
explicitly \cite{BeSiF1}. It has the curious property that it is
asymptotically
flat, but the fluid extends to spatial infinity.

Introducing the variable $v = (-V)^{1/2}$, the static field equations for
a perfect fluid with energy momentum tensor
\beq
T_{\mu\nu} = (\rho + p) u_\mu u_\nu+ p  g_{\mu\nu}
\eeq
with $u_\mu = v^{-1} \xi_\mu$ read
\beq
D^2 v = 4 \pi v(\rho + 3p)
\eeq
\beq
\R_{ij} = v^{-1} D_i D_j v + 4\pi(\rho - p)h_{ij}
\eeq
The asymptotic conditions (2.102,104) imply that $v \ra 1$ at infinity.
>From the
maximum principle for elliptic equations it follows that $0 < v < 1$ in
$N$. Since the surface of the star is at infinity for the Buchdahl
solution, the $v|_{\partial S}$, which is always equal to one in that
case, has to be replaced by the total mass $M$ (see Sect. 4.1).

Applying the contracted Bianchi identity to (5.12,13), there follows
\beq
D_i p = - v^{-1} (\rho + p) D_i v .
\eeq
Thus $p$ and $\rho$ are both functions of $v$ and
\beq
\frac{dp}{dv} = - v^{-1} (\rho + p) .
\eeq
Define the Cotton tensor of $h_{ij}$ $B_{ijk}$ by
\beq
B_{ijk} = 2D_{[k} \left(\R_{j]i} - \frac{1}{2}h_{j]i} \R\right).
\eeq
With the definition
\beq
W = (D_i v)(D^i v)
\eeq
the equations (5.12,13) now imply (see Lindblom \cite{LiFl}) that
\beqa
D^2W &=& \frac{1}{4} v^4 W^{-1} B_{ijk} B^{ijk} +
v^{-1} (D^i v)(D_i W) + 8\pi v(D^i v)(D_i \rho) + 
\frac{3}{4} W^{-1} (D^i W)(D_i W) \no \\
&& \mbox{} - 8\pi W(\rho + p) 
+ 16 \pi^2 v^2 (\rho + 3p)^2 - 4\pi v(\rho + 3p)
W^{-1} (D^iv)(D_i W).
\eeqa
In the spherically symmetric case $W = W_0$ has to be of the form
$W_0 = W_0(v)$. The ODE resulting in that case from (5.18), has, for
the equation of state (5.10), an explicit solution given by
\beq
W_0 = (1-v^2)^4 \left[ \frac{1}{16M^2} - \frac{\pi \rho_0}{3}
\left(\frac{1-v}{1+v}\right)^2\right].
\eeq
We assume that $\alpha = \frac{16\pi}{3}\; \rho_0 M^2 > 1$. The function
$W_0$ is defined for $v \in [0,1]$. It is positive for $v \in (v_c,1)$,
with $v_c = (\sqrt{\alpha} - 1)/(\sqrt{\alpha} + 1)$ and $W_0(v_c) = 0$,
$W_0(1) = 0$. Thus $W_0$ satisfies the correct boundary condition at the 
central value $v_c$ of $v$ and at infinity.

We now define, for the given solution $(v,h_{ij})$, the scalar function
\beq
\wt W - \wt W_0 = \left(\frac{1-v^2}{2}\right)^{-4}(W - W_0)
\eeq
and the conformally rescaled metric
\beq
\wt h_{ij} = v^{-2} \left(\frac{1-v^2}{2}\right)^4 h_{ij}.
\eeq
(The constant $M$ occurring in $W_0$ is taken to be the mass of the given
solution.) In the asymptotically flat, vacuum case discussed in Sect. 3.2 
one
finds that the metric $\wt h_{ij}$ extends smoothly to the manifold
$\wt N = N \cup \{\Lambda\}$, with $\Lambda$ the point at infinity.
This is also true for the Buchdahl solution, and we assume it to be true
for the given, a priori non-spherical one. After some calculations we
find that
\beq
\wt D^2(\wt W - \wt W_0) = \frac{1}{4} \wt W^4 \wt B_{ijk} \wt B^{ijk}
+ \frac{3}{4} \wt W^{-1} \wt D^i(\wt W - \wt W_0)\wt D_i(\wt W - \wt W_0).
\eeq
Since $\wt W,\wt W_0$ also extend smoothly to $\wt N$, the function
$\wt W - \wt W_0$ satisfies the elliptic equation with nonnegative
right-hand side on the compact manifold $\wt N$. After integrating (5.22)
over $\wt N$ (or by the maximum principle) it follows that $\wt B_{ijk}$
is zero and
\beq
\wt W = \wt W_0(v).
\eeq
It then follows from \cite{BeSiF2}, that the given model is
isometric to the Buchdahl solution with the same $\rho_0$ and the 
same $M$.

\subsection{Spherically Symmetric, Static Perfect Fluid Solutions }

The metric for a static spherically symmetric  spacetime can be  written
\beq
ds^2=-c^2e^{\nu(r)}dt^2 +e^{\lambda(r)} dr^2 + r^2(d\theta^2+\sin^2\theta\  d\phi^2)\ .
\eeq
For a derivation see \cite{HaEl,SchSph}. Here $c$ is a constant which plays
the role of the speed of light. In
Appendix B of \cite{HaEl} it is also demonstrated that the $r^2$ in
front of the sphere metric is
no loss of generality for a static perfect fluid with positive mass density and
pressure. Hence it
is impossible to have two centers or two infinities. The field equations for a
perfect 
fluid are
\beq
8\pi Gc^{-2}\rho r^2=e^{-\lambda}(r\lambda'-1)+1
\eeq
\beq
8\pi Gc^{-4}p r^2=e^{-\lambda}(r\nu'+1)-1
\eeq
\beq
8\pi Gc^{-4}p={1\over 2}e^{-\lambda}\left(\nu''+{1\over 2}\nu'^2+r^{-1}(\nu'-\lambda')-{1\over
2}\nu'\lambda'\right)
\eeq
A prime denotes a derivative with respect to $r$. We have written $-c^2\rho$ for the timelike
eigenvector of the energy--momentum tensor to make the comparison with the Newtonian equations
easier. The field equation imply `energy--momentum conservation', which is a single equation for a
static perfect fluid
\beq
2p'=-\nu'(p+c^2\rho)\ .
\eeq
The first exact solution of these equation was alredy found in 1918 by Karl
Schwarzschild, the
solution with constant density \cite{ExSol}.  We have three independent
ordinary differential
equations for for four functions. Hence, one function can be specified freely. The most physical
case is to prescribe an equation of state
$\rho=\rho(p)$. Equation (5.25) can easily be integrated:
\beq
e^{-\lambda}=1-{8\pi G\over c^2}{1\over r}\int r^2\rho(r)dr +{\rm const}\ .
\eeq
As we only are interested in solutions with a regular center of spherical symmetry we define
$\lambda$ as follows
\beq
e^{-\lambda}=1-{8\pi G\over c^2}{1\over r}\int_0^r r^2\rho(r)dr\ .
\eeq
The usual definition of the 'mass up to $r$', namely
\beq
m(r)=4\pi\int_0^r r^2\rho(s)ds
\eeq
gives 
\beq
e^{-\lambda}=1-{2 G\over c^2}{m(r)\over r}\ .
\eeq
It is also useful to introduce the following quantity which is related to the 'mean density up to
$r$'
\beq
w(r)= r^{-3}m(r)\ .
\eeq
Then (5.32) becomes 
\beq
e^{-\lambda}=1-{2 G\over c^2}r^2 w \ .
\eeq
Various forms of the equations (5.25-28) will be used. Equations (5.25),
(5.26) and
(5.28) contain all the information. If we eliminate $\nu'$ then (5.26)
and (5.28) imply the
Tolman -- Oppenheimer -- Volkoff equation \cite{Wald}
\beq
p'=-Gr\left(1-{2G\over c^2} r^2 w\right)^{-1}
\left({4\pi p\over c^2}+w \right)
\left({ p\over c^2}+\rho\right)
\eeq
If an equation of state is given we can integrate (5.28) 
\beq
\nu(r)=-\int_{p_0}^{p(r)} {2dp\over p+ c^2 \rho(p)} +{\rm constant}
\eeq
In this formula $p_0$ denotes the central pressure. If we add the definition
of $w$ then (5.35)
and (5.33) form an integro--differential system. Differentiating (5.33)
we obtain
\beq
w'={1\over r}(4\pi \rho-3w)
\eeq
In \cite{ReSch} the following theorem is proved. 

\paragraph{Theorem:} Let an equation of state $\rho(p)$ be given such that $\rho$ is defined for $p\ge0$,
non--negative and continuous for $p\ge0$, $C^\infty$ for $p>0$ and suppose that $d\rho/dp >0$ for
$p>0$.

Then there exists for any value of the central density $\rho_0$ a unique inextensible, static,
spherically symmetric solution of Einstein's field equation with a perfect fluid source and equation
of state $\rho(p)$. The matter either has finite extent, in which case a unique Schwarzschild
solution is joined on as an exterior field, or the matter occupies the whole space, with $\rho$
tending to $0$ as  $r$ tends to infinity.

There are two parts of the proof. The equations (5.35) and  (5.37) form
a system of ordinary
differential equations for $p(r),w(r)$. However, the system is singular at $r=$ and the first step
is to demonstrate that for each  value of the central density there is a unique solution such that
the spacetime is regular at the center. This is shown in \cite{ReSch}
or in \cite{Mak}. This
solution defines a neighborhood of a regular center and can be extended as long as $(1-{2G\over
c^2}r^2 w)$ remains positive. This can be seen as follows.  

Introduce the variables  first used by Buchdahl \cite{Bu1} 
\beq
y^2=1-{2G\over c^2}r^2 w\ , \ \ \ \   \zeta=e^{\nu/2}\ ,\ \ \ x=r^2
\eeq
Rewriting the equations in these variables and eliminating $p$ in 5.26)
and (5.28)  gives an
equation which is linear in $\zeta$ and $w$,
\beq
(1-{2G\over c^2}x w)\zeta_{,xx}-{G\over c^2}\zeta_{,x}(w+xw_{,x})_{,x}-{G\over 2c^2}w_{,x}\zeta=0
\eeq
or
\beq
(y\zeta_{,x})_{,x}-{G\over 2c^2}{w_{,x}\zeta\over y}=0
\eeq
Let $0\le x<x_0$ be an intervall such that $y^2=(1-{2G\over c^2}xw>0 $ and $p>0$. As the density
does not increase outwards we have $w_{,x}\le0$. Therefore 
\beq
(y\zeta_{,x})_{,x}\le0
\eeq
The equation (5.26) can be rewritten as 
\beq
y\zeta_{,x}={\zeta\over y}{G\over 2c^2}(w+{4\pi\over c^2}p)\ .
\eeq
>From (5.41) and (5.42) we obtain the inequality 
\beq
y\ge{w+4\pi p/c^2\over w_0+4\pi p_0/c^2}\ .
\eeq
Hence, we see that $y$ cannot vanish before $p$. 

Suppose $p(x_b)=0$. Then we call the corresponding $r_b$ the radius of the star. The Schwarzschild
solution is given in the form $e^{-\lambda}=e^{\nu}=1-A/r$ for some constant $A$. Hence, we
determine a unique exterior field by the condition $A=({2G\over c^2})m(r_b)$. In this way the
matter solution and the outside solution are joined only in a $C^0$--fashion because the boundary
density may be non--zero. If we introduce Gauss coordinates relative to the hypersurface $p=0$ the
metric is $C^1$. It is obvious that this metric cannot be extended because the area of the group
orbits $r={\rm constant}$ grows from $0$ to infinity.

Let us now consider the second possibility that $p(x)>0$ for all $x$. Because $p(x)$ is
monotonically decreasing for $x\to\infty $, $\lim_{x\to\infty}p(x)=p_\infty$ exists. This
implies that $p'$ tends to $0$ for $x\to\infty$. Since $y\le 1$, (5.35)
then implies $p_\infty=0$
and hence, using the equation of state, that $\rho\to 0$ as $x\to\infty$. As before the spacetime
is not extensible.
 
This completes our  outline of the proof. It shows in particular that for for $\rho(p)$ with
$\rho(0)=\rho_b>0$ the radius of the star has to be finite.

There are various exact global solutions  known. (For a useful list of such
solutions including a discussion of their physical acceptability has been
given by Delgaty and Lake \cite{DeL}.) For the 1--parameter
family of equations of state given by Equ. (5.10)  the whole
1--parameter family of solutions is known.  A 2--parameter family of
equations of state of interest for the  issue of section 3.1  is
investigated by Simon \cite{Si1}; all
the corresponding exact solutions are given.  

There are some conditions on the equation of state known, which allow to decide whether the radius
of the star is finite or infinite in the case of vanishing boundary density
$\rho_b$. In
\cite{ReSch} it is shown that the radius of the star is finite if
$\int_o^{p_0}dp/
\rho(p)^2$ is finite. Conversely,
$\int_o^{p_0}dp/ (\rho(p)c^{-2}p<\infty$ implies that the matter distribution is infinitely
extended. Both conditions depend only on the behaviour of the equation of state near the boundary
$p=0$.  Makino \cite{Mak} gives conditions for a finite radius in cases
which are not covered by the
above. He shows in particular, that for polytropic equations of state, $p=const.\rho^\gamma$ with
$4/3<\gamma<2$ the radius is finite.  

For finite distributions "Buchdahl's inequality"  holds \cite{Bu1}. 

\paragraph{Theorem:} For finite distributions with non--negative density and a monotonic equations of state
there holds
\beq
1-{2G\over c^2}{M\over r_b}>{1\over 9}\ .
\eeq

\paragraph{Proof:}
To obtain the inequality one compares the solution with a solution of constant density $\rho$, an
interior Schwarzschild solution. Equ. (5.40) implies for this solution
(written with an overbar) that
\beq
(\bar y\bar\zeta_{,x})_{,x}=0\ \ \Longrightarrow  \bar y\bar\zeta_{,x}=a={\rm constant} 
\eeq
We normalize $\zeta$ by the condition that at the boundary we have $\zeta_b=y_b$. Then we find
$a$ if we rewrite (5.26) in the new varables
\beq
{8\pi G\over c^2}p=4y^2{\zeta_{,x}\over \zeta}-{2G\over c^2}w
\eeq
and evaluate it at the boundary  as $a={2G\over c^2}\bar w$.

Then  (5.45) can be integrated with the result
\beq
\bar\zeta(x) ={1\over2}\left(1+2\bar\zeta(0) +\sqrt{1-{2G\over c^2}x\bar w} \right)
\eeq

Now (5.41) implies
\beq
 y\zeta_{,x}>( y\zeta_{,x})_b=\bar y\bar\zeta_{,x}
\eeq
 As $\bar y>y$  we obtain
\beq
\zeta_{,x}\ge \bar\zeta_{,x}={1\over2}\left(1+2\bar\zeta(0) +\sqrt{1-{2G\over c^2}x\bar w} \right)
\eeq
 As $\bar\zeta $ is positive we obtain at the boundary
\beq
y_b\ge -{1\over 2} y_b+{1\over 2}
\eeq
which is (5.44).

Buchdahl's inequality show that one can pack only a certain mass into a given fixed radius. The
physical reason is that the pressure is also a source of the gravitational field. In Newton's
theory there are constant density balls with a fixed radius for arbitrary density. In Einstein's
theory the central pressure diverges if the density approaches some maximum value. 

In \cite{BaRe} an analogue of Buchdahls inequality is derived  for
distributions in
which the the density is only assumed to be positive. There holds
$1-{2G\over c^2}{M\over r_b}>0$.   

Another important topic are bounds on the total mass of the system.  Suppose we know the
equation of state  only for $\rho<\rho_0$. Then we can estimate the mass and radius of a core in
which the density is greater $\rho_0$ as follows: Clearly, $m(r_0)>{4\pi\over 3}\rho_0(r_0)^3$; 
because of $y>0$ we have also $m(r_0)<{c^2\over 2G}r_0$. Hence the possible cores occupy a compact
part of the $m(r_0)$--$r_0$-- plane. Taking  intial values  from this part one can numerically
integrate outwards using the known equation of state, until the pressure vanishes. This was done in
\cite{Har}     for $\rho_0=5.1\times10^{14} g/cm^3$ and with a certain
realistic
equation of state for smaller densities.  All configurations  had a total mass smaller then  $5
M_\odot$. It is quite remarkable the the knowledge of the equation of state for a finite density
range allows to show such a bound on the total mass, assuming nothing but the monotonicity of the
equation of state in the unknown density range. This is not possible in Newton's theory. 

In the special case of bodies with a sharp edge, i.e $\rho_b>0$, we can
 combine the Buchdahl inequality (5.44) with the estimate $M=m(r_b)\geq 4\pi \rho_b r_b^3$ to
 obtain the mass bound 
\beq
 M\leq\left(\frac{2}{3}\right)^3\left(\frac{3c^6}{4\pi G^3\rho_b}\right)^{1/2}\ .
\eeq

Let us finally compare with Newton's theory. In (5.35) it is almost
obvious that for
$c\to\infty$  one obtains the Newtonian equation for the pressure. The relativistic corrections
show how "the pressure enters in the active and passive gravitational mass". The first factor
describes an effect of the geometry. Static fluid ball are the simplest examples of families of
relativistic solutions with a Newtonian limit \cite{EhN}. 

\subsection{Spherically Symmetric, Static Einstein--Vlasov Solutions }

In recent years  existence and further properties of solutions of
Einstein's field equations for a collisionless gas have been shown
\cite{ReRe}. 
The Vlasov--Einstein-- system determines the spacetime metric and the
distribution function $f(x^\mu,p^\mu)$ describing the particles.
$$
p^\mu\partial_{x^\mu}f -\Gamma^\mu_{\nu\sigma}p^\nu p^\sigma
\partial_{p^\mu} f=0
$$
\beq
T^{\mu\nu}=\int p^\mu p^\nu |g|^{1/2}{d^4p\over m}
\eeq
$$
G^{\mu\nu}=8\pi T^{\mu\nu}.
$$
 In the static spherically symmetric case and for the metric (5.24), these
equations
reduce to (r=\ $|x^i|, v^i$  are the spatial frame components of $p^\alpha$)
\beq
{v^i\over \sqrt{1+v^2}}\partial_{x^i}f - \sqrt{1+v^2}\nu'{x^i\over r}\partial_{v^i}f=0
\eeq
\beq
8\pi Gc^{-2}\rho r^2=e^{-\lambda}(r\lambda'-1)+1
\eeq
\beq
8\pi Gc^{-4}p r^2=e^{-\lambda}(r\nu'+1)-1
\eeq
where
\beq
\rho(x)=\rho(r)=\int_{R^3}f(x^i,v^i)\sqrt{1+v^2}dv\ ,
\eeq
\beq
p(x)=p(r)=\int_{R^3}f(x^i,v^i)\left({x^iv_i\over r}\right){dv\over \sqrt{1+v^2}}\ .
\eeq
The distribution function is assumed to be spherically symmetric.

Rein and Rendall \cite{ReRe} show the existence of asymptotically flat
solutions, regular at
the center, with finite total mass and finite  extension of the matter and isotropic
pressure. It is also possible to construct solutions with anisotropic pressure;
Furthermore  shells of finite extent of matter around a regular center or a black
hole can be constructed \cite{Rein}.

\clearpage
\addcontentsline{toc}{section}{Index}
\flushbottom
\printindex

\end{document}